\UseRawInputEncoding
\documentclass[aps,twocolumn,pra,superscript,floatfix,superscriptaddress,showpacs,footinbib,reprint]{revtex4-1}
\usepackage{amssymb}

\usepackage[pdftex]{graphicx}
\usepackage{dcolumn}
\usepackage{bm}
\usepackage{amsmath}
\usepackage{nicefrac}
\usepackage{array}
\usepackage{color}
\usepackage{float}
\usepackage{subfigure}
\usepackage{dsfont}
\usepackage{txfonts}
\usepackage{wasysym}
\usepackage{multirow}
\usepackage{sidecap}
\usepackage{xcolor,cancel}
\usepackage{verbatim}
\usepackage[plainpages=false,pdfpagelabels,colorlinks=true,linkcolor=red,urlcolor=blue,
citecolor=blue,pdftitle={new-version},pdfauthor={},pdfdisplaydoctitle=true]{hyperref}
\usepackage[normalem]{ulem}

\begin{document}

\title{Ferromagnetism in armchair graphene nanoribbon heterostructures}

\author{P.~A.~Almeida}
\affiliation{Instituto de F\'isica, Universidade Federal de Uberl\^andia, 
Uberl\^andia, Minas Gerais 38400-902, Brazil.}

\author{L.~S.~Sousa}
\affiliation{Instituto de F\'isica, Universidade Federal de Uberl\^andia, 
Uberl\^andia, Minas Gerais 38400-902, Brazil.}

\author{Tome~M.~Schmidt}
\affiliation{Instituto de F\'isica, Universidade Federal de Uberl\^andia, 
Uberl\^andia, Minas Gerais 38400-902, Brazil.}

\author{G.~B.~Martins}
\affiliation{Instituto de F\'isica, Universidade Federal de Uberl\^andia, 
Uberl\^andia, Minas Gerais 38400-902, Brazil.}
\email[Corresponding author: ]{gbmartins@ufu.br}

\date{\today}
\begin{abstract}
	We study the properties of flat-bands that appear in a heterostructure composed of strands of 
	different widths of graphene armchair nanoribbons. One of the flat-bands is reminiscent 
	of the one that appears in pristine armchair nanoribbons and has its origin in a 
	quantum mechanical destructive interference effect, dubbed `Wannier orbital states' 
	by Lin \emph{et al.} in Phys. Rev. B 79, 035405 (2009). The additional flat-bands found 
	in these heterostructures, some reasonably closer to the Fermi level, seem to be generated by a 
	similar interference process. 
	After doing a thorough tight-binding analysis of the band structures of the different kinds of 
	heterostructures, focusing in the properties of the flat-bands, we use Density Functional Theory 
	to study the possibility of magnetic ground states when placing, through doping, the Fermi energy 
	close to the different flat-bands. Our DFT results confirmed the expectation that these 
	heterostructures, after being appropriately hole-doped, develop a ferromagnetic ground state 
	that seems to require, as in the case of pristine armchair nanoribbons, the presence of 
	a dispersive band crossing the flat-band. In addition, we found a 
	remarkable agreement between the tight-binding and DFT results for the charge density distribution 
	of the so-called Wannier orbital states. 
\end{abstract}

\maketitle

\section{Introduction} \label{sec:intro}
Strong correlations in magic-angle twisted 
bilayer graphene (TBG), discovered in 2018~\cite{Cao2018} (see Ref.~\cite{Eva2020} for a review), 
were associated to the presence of strongly correlated states in flat mini-bands 
of the hexagonal Moir\'e superlattice, as previously predicted by band 
structure calculations~\cite{Santos2007,Morell2010,Bistritzer2011}. 
Recently, ARPES measurements~\cite{Lisi2021} have provided direct evidence for the 
existence of flat-bands in magic-angle TBG. These developments have greatly increased the interest in the 
study of low-dimensional systems presenting bands with zero (or quasi-zero) 
dispersion. 

Indeed, in the last one year alone, there has been new flat-band research in 
many different areas, like their experimental observation in atomically precise 
one-dimensional (1D) chains~\cite{Huda2020}, 
as well as the study of flat-bands in strongly correlated 
systems~\cite{Liu2020,Leite2021,chan2021,Orito2021,richter2021,kumar2021,cadez2021,mizoguchi2021a,chung2021}, 
search for flat-bands in kagome-type lattices~\cite{Meier2020,Ye2021}, study of symmetry aspects of 
flat-band systems~\cite{Morfonios2021,Rhim2019,Hwang2021}, holographic construction of flat-bands~\cite{Grandi2021}, 
flat-bands in pyrochlore lattices~\cite{nakai2021,Mizoguchi2021}, 
analysis of randomness in flat-band Hamiltonians~\cite{hatsugai2021}, 
topological aspects of flat-band systems~\cite{wang2021,zurita2021,kruchkov2021,liu2021,calugaru2021,luo2021}, 
construction of flat-band tight-binding models starting from 
compact localized states~\cite{Graf2021}, and study of flat-bands in graphene and graphene-like 
lattices~\cite{Li2021,bao2021,wang2021a,pathak2021,desousa2021}. 

For a brief review of the research in flat-bands, describing initial theoretical 
proposals in the late 1980s~\cite{Sutherland1986,Lieb1989}, their association to 
topological phases~\cite{Parameswaran2013,Bergholtz2013}, and their 
possible realization in superconducting wire networks, cold atoms in optical lattices, 
and photonic systems, see Ref.~\cite{Leykam2018}. For a description of strongly correlated 
ground states associated to dispersionless bands, see Ref.~\cite{Derzhko2015}.

Following the development of a bottom-up procedure for atomically precise synthesis of 
semiconducting graphene nanoribbons (GNRs) with different width, edge, and end
termination~\cite{Cai2010}, a seminal paper by Steven Louie's group in 2017~\cite{Cao2017} showed that 
these synthesized armchair GNRs (AGNRs) strands belonged to different 
topological phases, protected by spatial symmetries 
and with a $Z_2$ topological invariant whose value was dictated by their width and terminating unit cell. 
Thus, the bulk-boundary correspondence principle~\cite{Konig2008,Hasan2010,
Qi2011,Yan2012,Ren2016,Bansil2016} imposes that at the interface between two finite AGNRs, 
with different $Z_2$ values, a topologically protected localized state 
should exist, with its energy located inside the AGNR gap. This 
expectation was confirmed by Density Functional Theory (DFT) calculations~\cite{Cao2017}. 
The following year, two experimental groups, one in Europe~\cite{Rizzo2018} 
and the other in the USA~\cite{Groning2018}, published side-by-side Nature 
papers presenting DFT and tight-binding simulations of Scanning Tunneling Spectroscopy (STS) 
measurements in superlattices of short AGNR strands, alternating between finite and vanishing $Z_2$ values, that 
indicated the presence, inside the (overall) AGNR gap, of a dimerized chain 
band structure. A Su-Schrieffer-Heeger (SSH) 
effective model (initially proposed to describe polyacetylene~\cite{Su1979}, and recently revived 
as a prototypical model for a one-dimensional topological insulator~\cite{Asboth}), 
was shown to qualitatively describe the 
experimental results. Thus, in what was described as a \emph{hierarchically engineered 
one-dimensional topological system}~\cite{Rizzo2018}, the AGNR heterostructure, with 
topologically non-trivial properties 
(i.e., a topologically protected end state), is itself composed of alternating topologically-trivial and non-trivial 
building blocks. Besides the ability of considerably decreasing the AGNR's spectral gap (with the 
recent observation of metallicity in an AGNR heterostructure~\cite{Rizzo2020}--notice that all  
AGNRs are  actually semiconducting~\cite{Son2006}), the properties of 
these heterostructures, as implied by the results 
presented in Refs.~\cite{Groning2018,Rizzo2018}, have generated much attention, as they represent one 
of the first stable materials (besides polyacetylene) that simulates the SSH model, 
which up to now had been simulated mainly in cold-atom~\cite{Meier2016}, 
engineered atomic lattices~\cite{Drost2016,Yan2019}, 
photonic~\cite{saxena2021}, acoustic~\cite{Coutant2021}, and 
mechanical~\cite{Chen2014,Huber2016} experimental configurations. 
Very recent work, extending the results in Refs.~\cite{Groning2018,Rizzo2018}, 
may be found in Refs.~\cite{Sun2021,Li2021a}.

\begin{figure}
\includegraphics[trim=1.6cm 2cm 1.6cm 2cm, clip=true,width=0.45\textwidth]{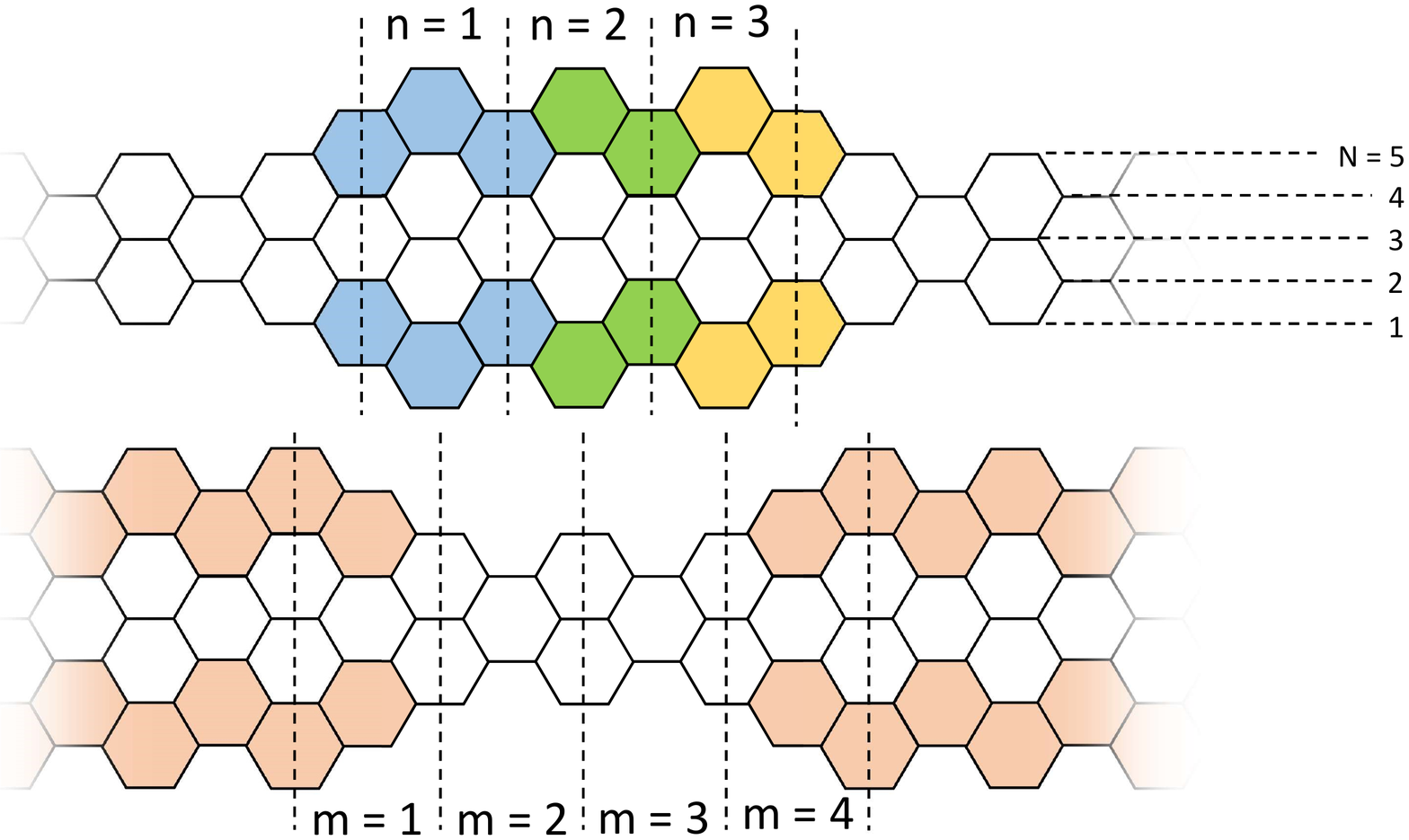}
\caption{Schematic representation of the meaning of the parameters 
	$N$, $n$, and $m$ in an N-AGNR(n,m) heterostructure. See text for details. 
}
\label{fig1}
\end{figure}

\begin{figure}
\includegraphics[width=0.45\textwidth,angle=0.]{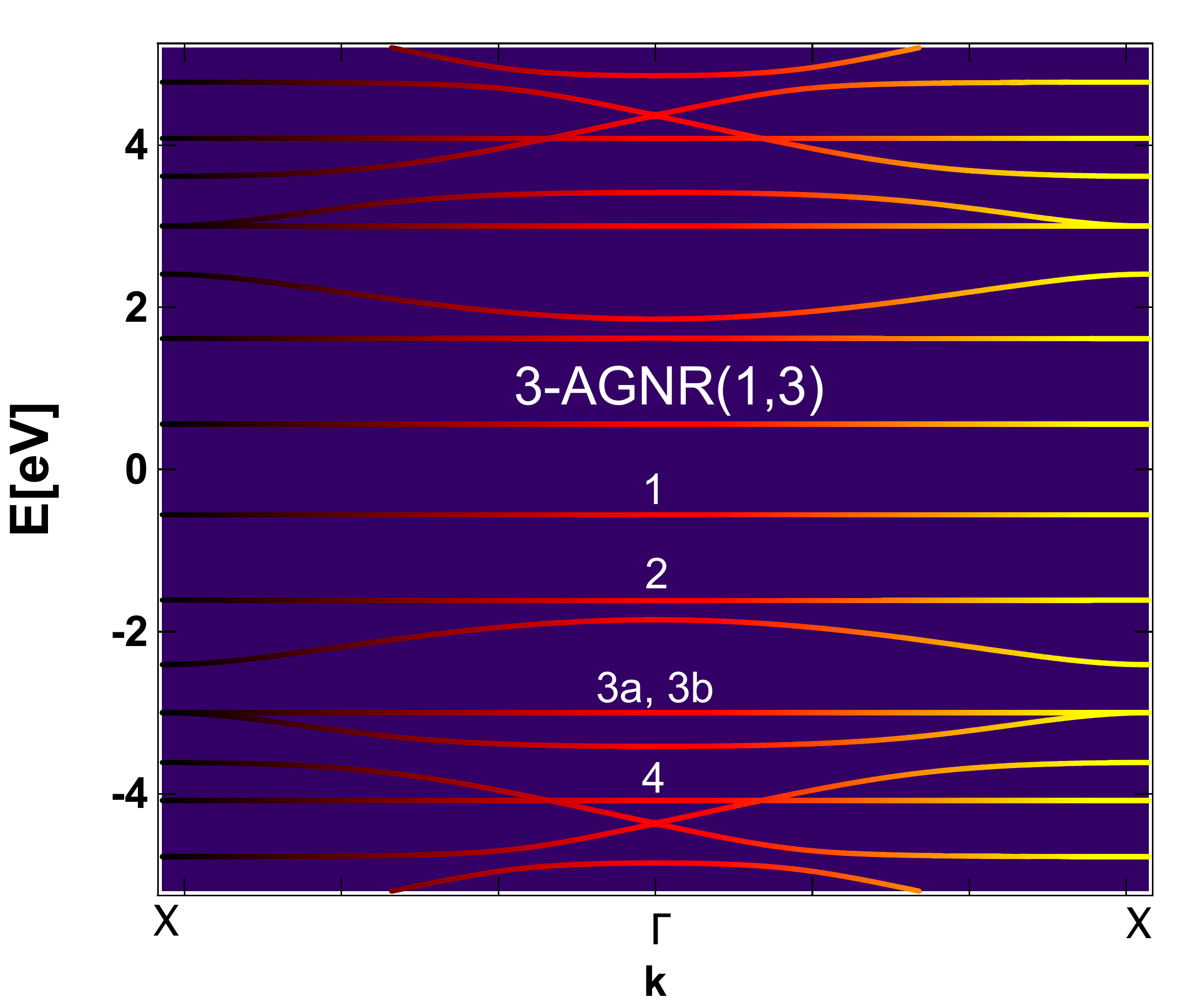}
\caption{Tight-binding band structure of a 3-AGNR(1,3) heterostructure. 
	The flat-bands are indicated by labels $1$, $2$, $3a$, $3b$, and $4$, 
	starting from the Fermi energy at half-filling ($E=0$). Note that 
	band $3$ is double-degenerate. 
}
\label{fig2}
\end{figure}

A much less studied aspect of these AGNR heterostructures is the presence of dispersionless bands in their 
band structure. In this work, using the tight-binding method and DFT, we 
systematically analyze how the presence or not of flat-bands, their proximity to the Fermi energy, 
their interplay with nearby dispersive bands, as well as if they give origin or not to a ferromagnetic 
ground state, depends on the parameters that define the AGNR heterostructure. Our results show that,
indeed, the majority of the heterostructures studied through tight-binding present 
several flat-bands that can be associated to `Wannier orbital' 
states, as formerly seen in pristine AGNRs~\cite{Lin2009}. By appropriately hole-doping 
these heterostructures, i.e., bringing the Fermi energy close to a flat-band, a ferromagnetic 
ground-state is observed through DFT simulations. The ferromagnetic exchange coupling 
at the flat-band appears to be mediated by a dispersive band that crosses it~\cite{Lin2009}.

Before presenting the organization of the paper, we want to emphasize 
that we do not investigate the topological properties of the heterostructures 
studied here. We just, eventually, point out some possible connections between ferromagnetism 
and the SSH effective model, which may motivate further research on that.

The organization of the paper is as follows: In Sec. \ref{sec:model} we introduce the AGNR heterostructure 
parameters $N$, $n$, and $m$, together with the tight-biding Hamiltonian that models it, while in 
Sec.~\ref{sec:N=3/n=1/m=3}, to illustrate the appearance of flat-bands in 
these heterostructures, we present the tight-binding results for heterostructures with the second smallest 
unit cell, i.e., $N=3$, $n=1$, and $m=3$, showing the presence of four valence flat-bands 
(with respective particle-hole symmetric conduction band partners). 
Then, in Sec.~\ref{sec:vary-N}, we show that 
flat-bands survive for `backbones' $N=5$ and $N=7$, and also present the profile of the 
`Wannier orbital' states associated to each one of the four lowest energy flat-bands. 
For $N=9$, the flat-bands present for smaller values of $N$ acquire dispersion. 
In Sec.~\ref{sec:n-and-m}, we keep $N=3$ and vary the other two parameters, $n$ and $m$, 
and analyze their influence over the flat-bands and the corresponding 
`Wannier orbital' states (which, from now on, will be called Wannier-like states). 
This will set the stage for an \emph{ab-initio} DFT analysis 
of the ferromagnetic ground state present for varying hole-doping in Sec.~\ref{sec:DFT}. 
Finally, in Sec~\ref{sec:sum$conc}, we present a summary of the results obtained and our concluding remarks. 
For completeness sake, in Sec.~\ref{ap:seca1} of the Appendix, we present the Hamiltonian 
in real and reciprocal space for an N-AGNR(1,3) heterostructure (see next Section, for notation). 
In addition, in Secs.~\ref{ap:seca2} and \ref{ap:seca3} of the Appendix, we discuss 
the effects of adding a next-nearest-neighbor (NNN) hopping to the main-text tight-binding 
calculations and briefly present tight-binding and DFT band structures for an alternative 
(less symmetric) type of heterostructure that has also been synthesized in the 
laboratory~\cite{Groning2018,Rizzo2018}.  

\section{Model for the heterostructures}\label{sec:model}
\subsection{The geometry of the N-AGNR(n,m) heterostructures}
In Ref.~\cite{Groning2018}, two types of AGNR heterostructures were introduced, 
the so-called `inline' and `staggered' heterostructures. In this paper, we will analyze the properties 
just of inline heterostructures (which we will name N-AGNR(n,m) heterostructures), 
since they present more flat-bands than the staggered heterostructures. 

\begin{figure}
 \begin{minipage}{.5\textwidth}
 \includegraphics[trim=7cm 0cm 8cm 0cm,clip=true,width=\textwidth]{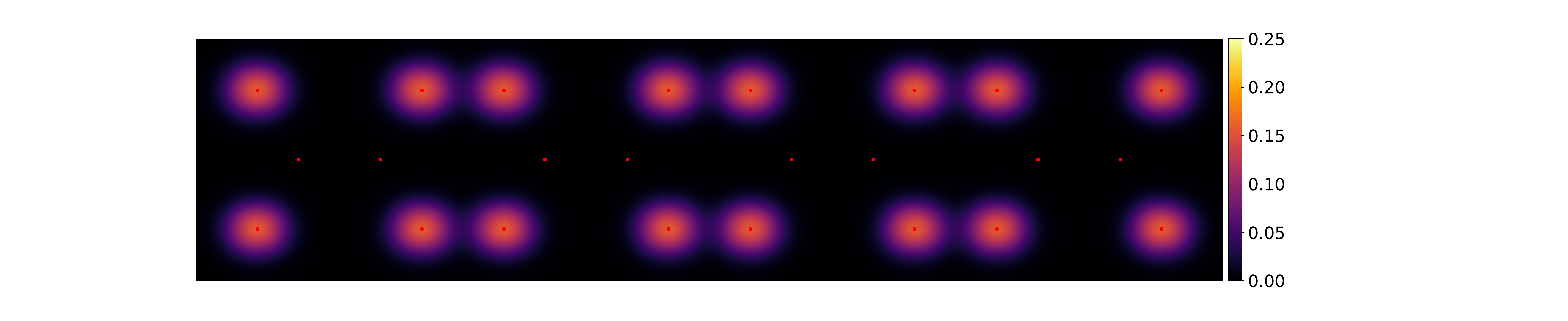}
 \end{minipage}
\caption{Charge density of the Wannier-like state for the $-t$ flat-band on a pristine 3-AGNR. 
}
\label{fig3}
\end{figure}

In Fig.~\ref{fig1}, we schematically show how the unit cell of an N-AGNR(n,m) heterostructure 
is built. In the top panel, the parameter $n$ indicates how many adjacent unit cells 
(delimited by vertical dashed lines)
of the so-called backbone (a pristine N-AGNR, depicted in white), containing $N=5$ dimers in each unit cell, 
as indicated in the right, will be extended into unit cells containing $N+4$ dimers. 
As indicated in Fig.~\ref{fig1}, this is done, for the first of the $n$ unit cells, by adding 
six carbons to the top and bottom of the unit cell. This adds three extra 
benzene rings, colored in cyan, to the top and bottom of the unit cell.  To extend the 
next unit cell (adjacent to the right), just four extra carbon atoms are needed to add two 
more benzene rings, colored in green. This second step is repeated until all $n$ 
adjacent unit cells are extended. The top panel in Fig.~\ref{fig1} shows the result for $n=3$. 
Finally, in the bottom panel, $m$ indicates how many unit cells away from the last extended unit cell 
we will repeat the process of extending $n$ unit cells. There is an important detail here: 
we count $m$ from the \emph{center} of the last extended unit cell to the center 
of the first extended unit cell of the next $n$-group to the right (notice the positioning 
of the vertical dashed lines in the bottom panel, see Fig.~S2 in 
Ref.~\cite{Groning2018}). Therefore, the unit cell of the N-AGNR(n,m) heterostructure 
thus obtained will contain $n+m-1$ unit cells of the original backbone. It is clear that 
$m \geq 2$, since $m=1$ produces an uniform AGNR with a width equal to $N+4$. 

\begin{figure}
 \begin{minipage}{.5\textwidth}
  \includegraphics[trim=7cm 1.2cm 8cm 1.3cm, clip=true, width=\textwidth]{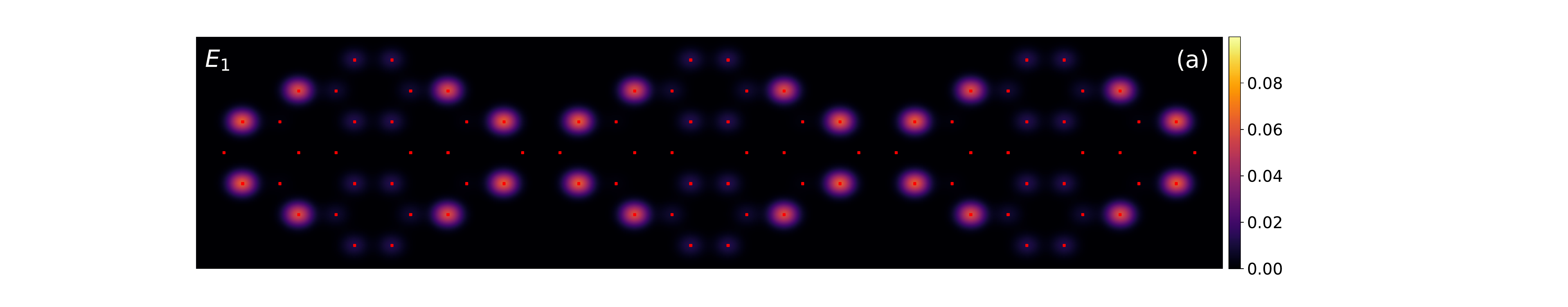}
 \end{minipage}
 \begin{minipage}{.5\textwidth}
  \includegraphics[trim=7cm 1.2cm 8cm 1.3cm, clip=true, width=\textwidth]{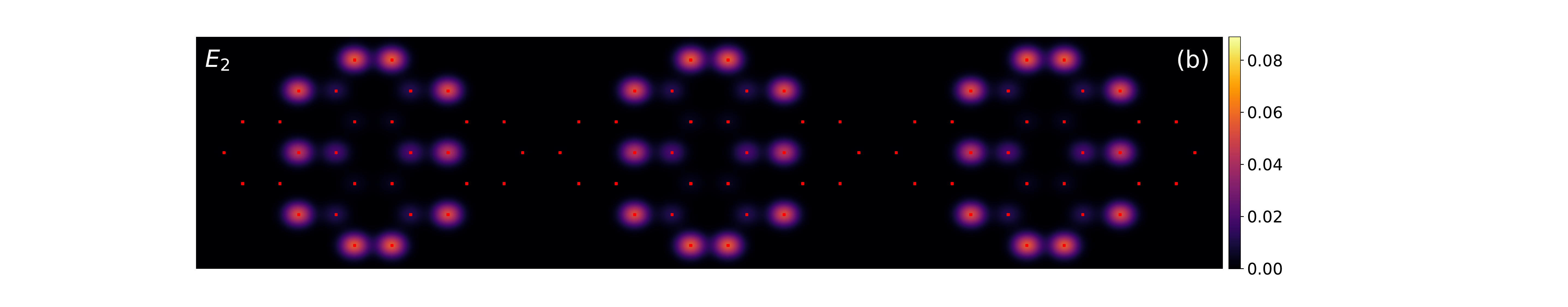}
 \end{minipage}
\begin{minipage}{.5\textwidth}
  \includegraphics[trim=7cm 1.2cm 8cm 1.3cm, clip=true, width=\textwidth]{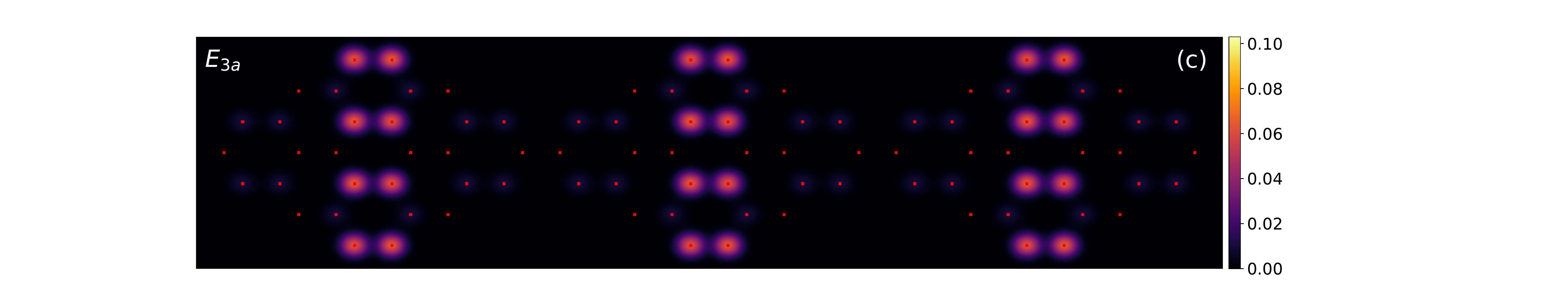}
 \end{minipage}
 \begin{minipage}{.5\textwidth}
  \includegraphics[trim=7cm 1.2cm 8cm 1.3cm, clip=true, width=\textwidth]{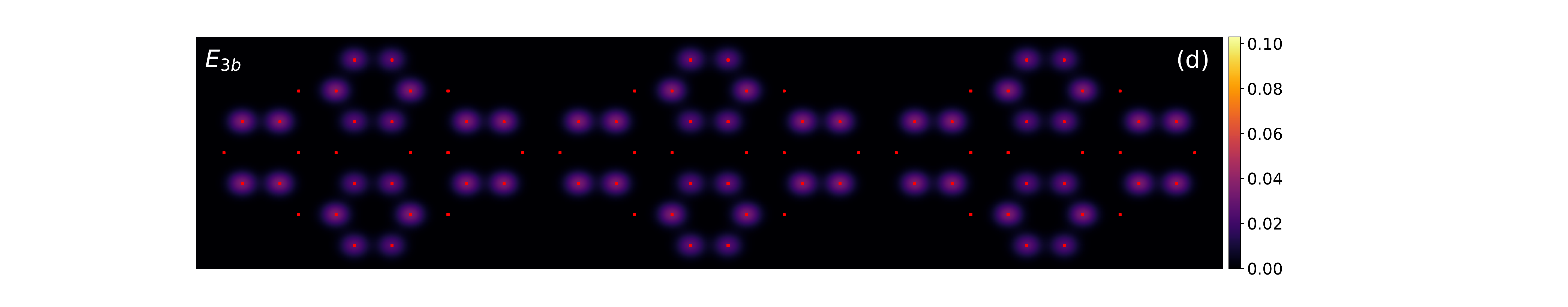}
 \end{minipage}
  \begin{minipage}{.5\textwidth}
  \includegraphics[trim=7cm 1.2cm 8cm 1.3cm, clip=true, width=\textwidth]{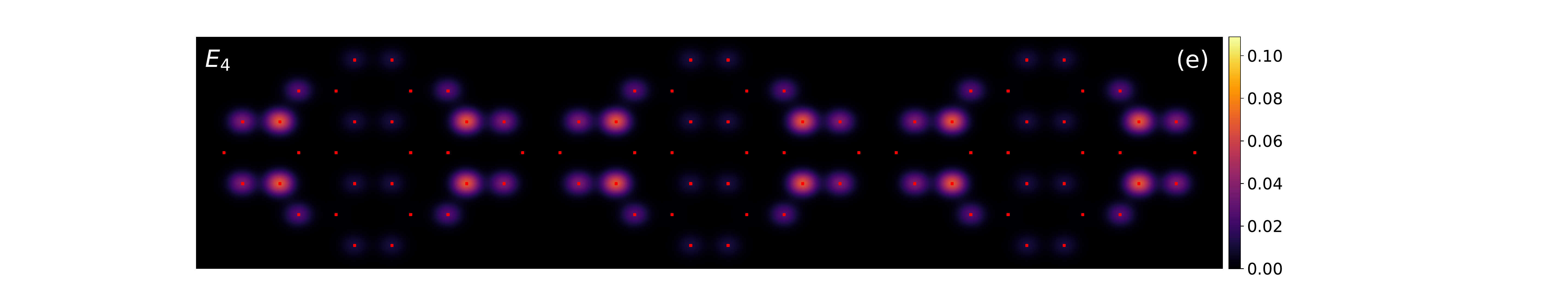}
 \end{minipage}
\caption{Wannier-like states for all four flat-bands 
in a 3-AGNR(1,3). From top to bottom, corresponding band energies 
are $E_1=-0.56$, $E_2=-1.61$, $E_{3a}=E_{3b}=-3.00$, and $E_4=-4.08$~eV. 
	}\label{fig4}
\end{figure}

\subsection{Tight-binding Hamiltonian}

The band structure of these N-AGNR(n,m) heterostructures 
will be simulated using a tight-binding Hamiltonian 

\begin{eqnarray}
	H_{\mathrm{tb}} &=&-t\sum_{\left\langle i,j\right\rangle \sigma}c_{i\sigma}^{\dagger }c_{j\sigma}, 
\label{eq:Hamilt1}
\end{eqnarray}

\noindent where $c_{i\sigma}^{\dagger }$ ($c_{i\sigma}$) creates (annihilates) 
an electron in site $i$ with spin $\sigma$ and $\left\langle i,j \right\rangle $ 
runs over nearest-neighbor sites.  This Hamiltonian describes nearest-neighbor hoppings 
with transfer integral $t$, where a typical value 
found in the literature for this parameter is 
$t \sim 3.0$~eV~\cite{Neto2009}. In Appendix \ref{ap:seca1}, a specific expression 
will be given for Eq.~\eqref{eq:Hamilt1} for a 3-AGNR(1,3) 
heterostructure, in real and reciprocal space. 

In Sec.~\ref{sec:DFT}, long-range Coulomb interactions 
will be added within the DFT framework. A hybrid
functional for the exchange-correlation term will be included in the DFT
to better describe the Coulomb interactions as well as the Wannier-like states.
The calculation methodology will be detailed in Sec.~\ref{sec:DFT} as well.

\begin{figure}
 \begin{minipage}{.5\textwidth}
 \includegraphics[width=\textwidth]{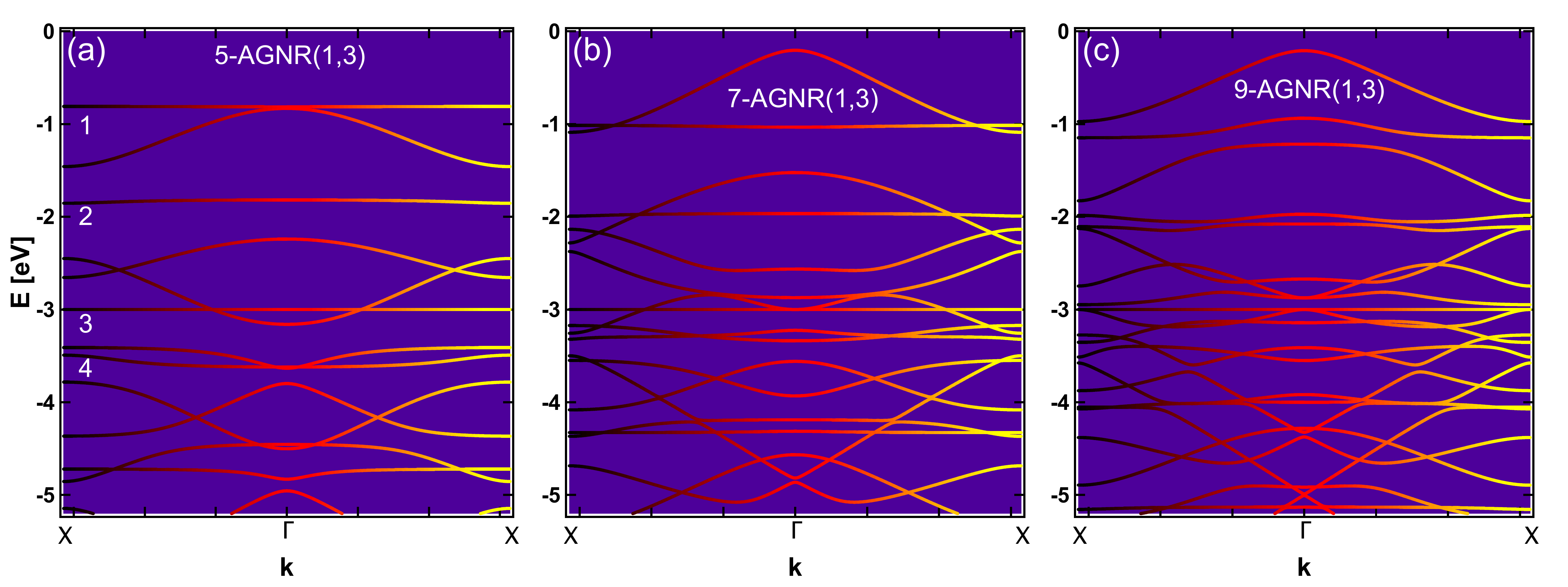}
 \end{minipage}
	\caption{Band structures for N-AGNR(1,3) heterostructures for 
	$N=5$, $7$, and $9$ in panels (a), (b) and (c), respectively. 
	Although it is not so apparent, for $N=9$ the only flat-bands 
	left is the pair $\pm t$. Aside from the $-t$ flat-band, the other three 
	flat-bands in panels (a) and (b) have changed their positions 
	in relation to the $N=3$ results (see Fig.~\ref{fig2}). 
	Note the scale, with only negative energies, to improve 
	readability. The band labeled $4$ in panel (a) has acquired 
	dispersion (compare to the corresponding band in Fig.~\ref{fig2}.)}
\label{fig5}
\end{figure}

In the next section, we will present tight-binding results 
for the band structure of a 3-AGNR(1,3) heterostructure. 
Note that the tight-binding and DFT band structures will be given in units of eV. 

\section{Flat-bands for a 3-AGNR(1,3) heterostructure}\label{sec:N=3/n=1/m=3}

In Fig.~\ref{fig2}, we show the tight-binding band structure for a 3-AGNR(1,3) 
heterostructure, for $t=3.00$~eV (the nearest-neighbor hopping integral value 
we will use for all tight-binding calculations).  
For the energy-interval shown, we label the negative energy flat-bands 
as $1$, $2$, $3a$, $3b$, and $4$, starting from the closest 
one to the Fermi energy (at half-filling). Their respective energies are $E_1=-0.56$, $E_2=-1.61$, 
$E_{3a}=E_{3b}=-3.00=-t$, and $E_4=- 4.08$~eV, where the band at 
$-t$ is \emph{double-degenerate}. 

\begin{figure}
 \begin{minipage}{.5\textwidth}
  \includegraphics[trim=7cm 1.8cm 8cm 1.3cm, clip=true, width=\textwidth]{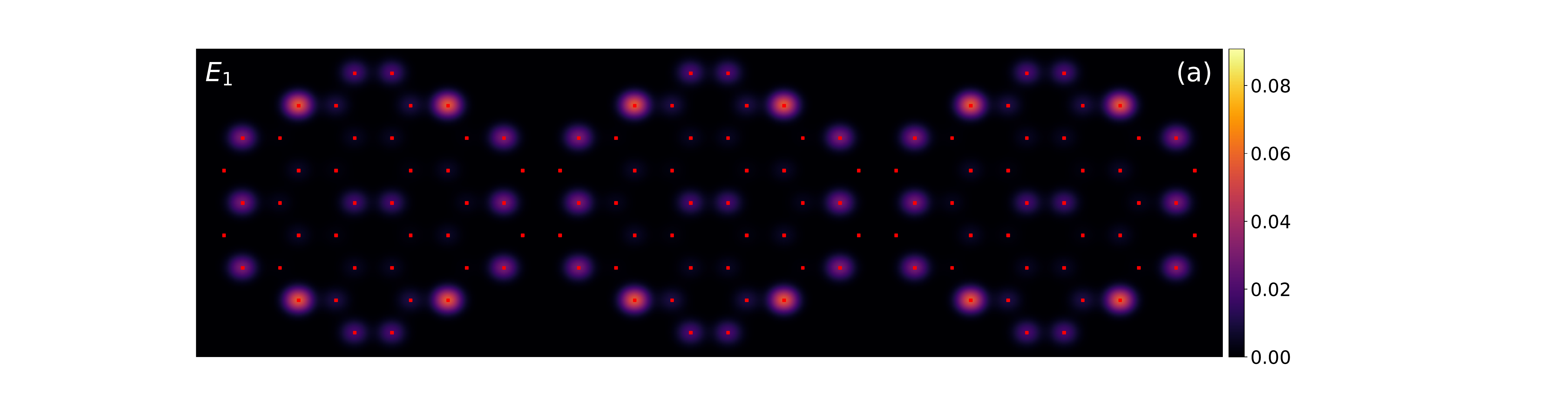}
 \end{minipage}
 \begin{minipage}{.5\textwidth}
  \includegraphics[trim=7cm 1.8cm 8cm 1.3cm, clip=true, width=\textwidth]{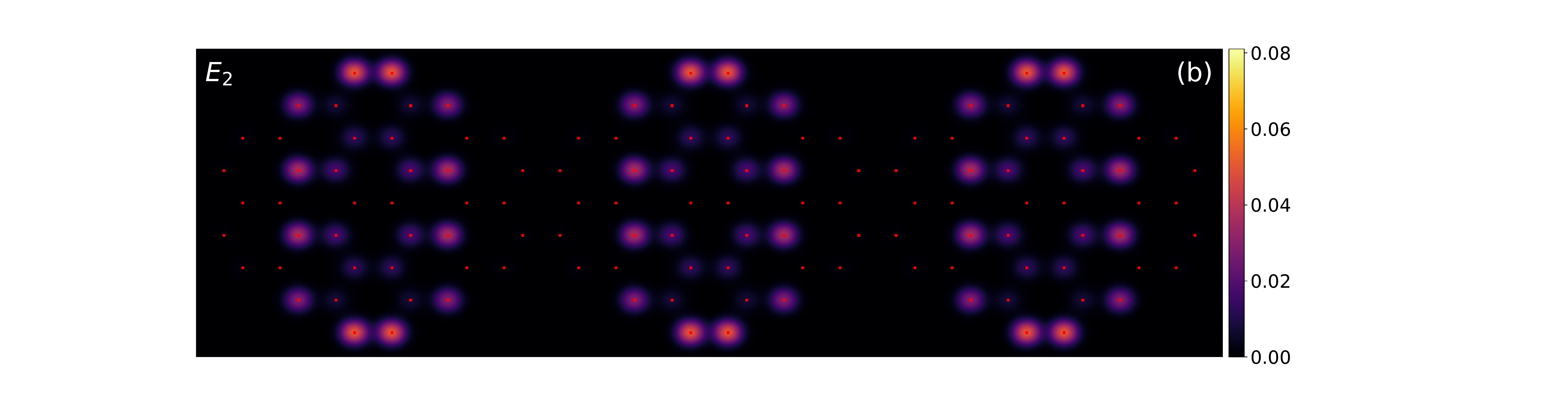}
 \end{minipage}
 \begin{minipage}{.5\textwidth}
  \includegraphics[trim=7cm 1.8cm 8cm 1.3cm, clip=true, width=\textwidth]{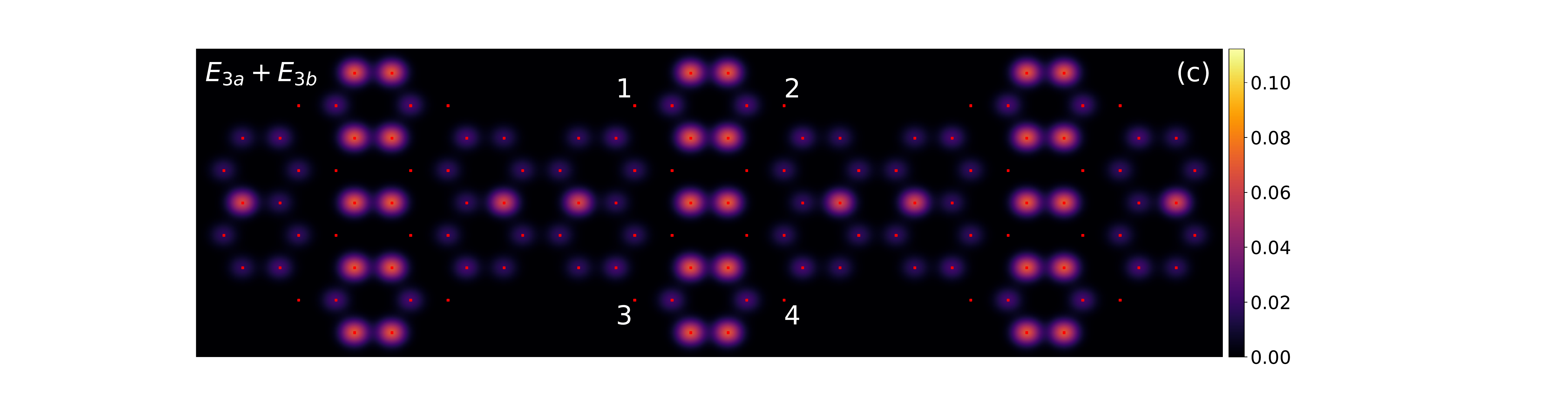}
 \end{minipage}
\caption{Wannier-like states for three flat-bands in a 5-AGNR(1,3). Numbers 1 to 4 
	in panel (c) indicate destructive quantum interference sites that prevent a continuous 
	nearest-neighbor path from existing, which would connect all unit cells across 
	the heterostructure, rendering state $E_{3b}$ dispersive. 
	}\label{fig6}
\end{figure}

It is relatively well known~\cite{Lin2009} that N-AGNRs (pristine, with no extensions) 
with odd-N present two perfectly flat-bands 
at $\pm t$, and Fig.~\ref{fig2} shows that this also happens 
for the 3-AGNR(1,3) heterostructure (energy $E_{3a}=E_{3b}=-t$). 
As a matter of fact, this is true for all odd-N N-AGNR(1,3) 
heterostructures we have investigated, with the difference 
that for $N=3$, $5$ and $7$ there are additional flat-bands at higher and lower energies, 
as shown in Fig.~\ref{fig2}. For $N \geq 9$, these additional flat-bands acquire dispersion 
(see Sec.~\ref{sec:vary-N}). One interesting point is that, in the N-AGNR(1,3) heterostructures, 
the $\pm t$ bands are double-degenerate for $N=3$ and $5$, however, this degeneracy is lifted for 
$N \geq 7$ (see Sec.~\ref{sec:vary-N}). 

\subsection{The Wannier-like states}

In Ref.~\cite{Lin2009}, a very interesting analysis is done of the magnetism of these 
$\pm t$ flat-bands that are present in the odd-N AGNR (without extensions, i.e., 
pristine AGNR). Indeed, the origin of the zero-dispersion is that the Bloch states 
associated to the $\pm t$ bands are formed by `isolated' clusters of charge inside 
each unit cell (the so-called `Wannier orbital' states, or Wannier-like states), which have 
\emph{zero} overlap with the clusters in adjacent unit cells. 
This happens because of destructive quantum interference~\cite{Lin2009}. 
This phenomenon is shown in Fig.~\ref{fig3}, which shows the integrated charge density 
(over all $k$-values) for $E=-t$ in each site of an $N=3$ pristine AGNR. 
Figure \ref{fig3} simulates the local density of states (LDOS) an Scanning 
Tunneling Microscope tip would observe in case its parameters were set to capture 
just the $E=-t$ states of a 3-AGNR. It is remarkable that each and every one of the 
different Bloch states (for different $k$-values in the Brillouin zone) at $E=-t$ 
has the same LDOS profile as the one shown in Fig.~\ref{fig3} (see Ref.~\cite{Lin2009} 
for details). It is worth mentioning that these so-called Wannier-like states are also called 
`compact localized states'~\cite{Morfonios2021}, which, as shown in Fig.~\ref{fig3}, are localized on a subset 
of lattice sites, with zero amplitude in the rest of the lattice. 
As shown in the Introduction, they have recently 
attracted a great deal of attention. A discussion of their properties and the relevant 
literature may be found in Ref.~\cite{Morfonios2021}. 

\begin{figure}
 \begin{minipage}{.5\textwidth}
  \includegraphics[trim=7cm 2.0cm 8cm 3.1cm, clip=true, width=\textwidth]{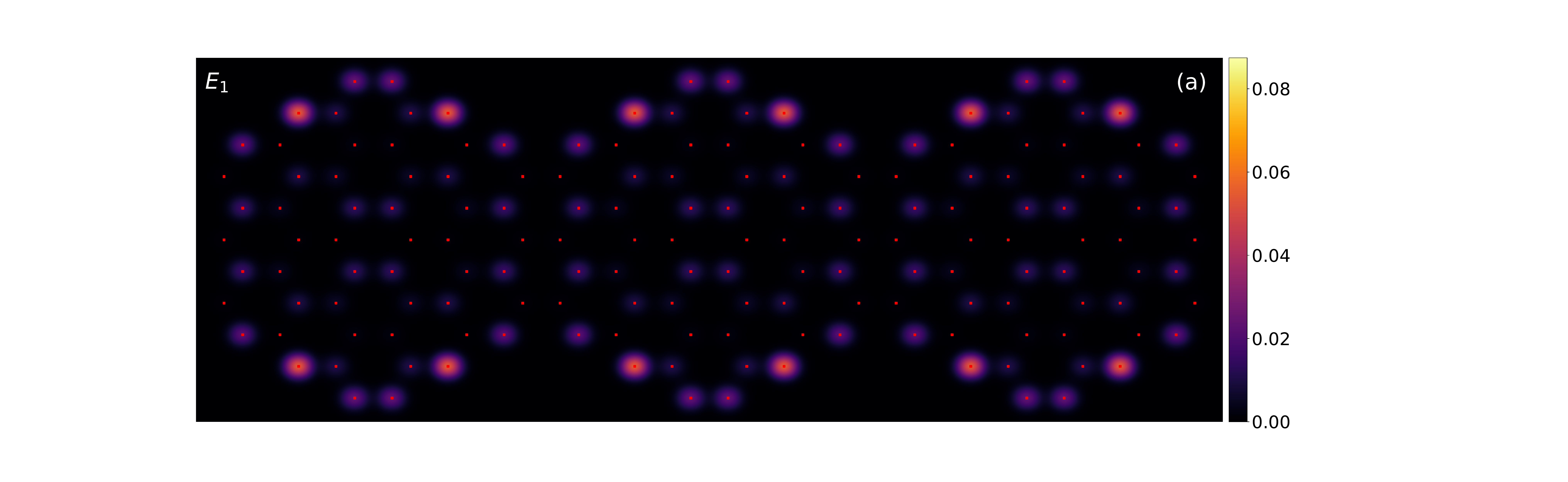}
 \end{minipage}
 \begin{minipage}{.5\textwidth}
  \includegraphics[trim=7cm 2.0cm 8cm 3.1cm, clip=true, width=\textwidth]{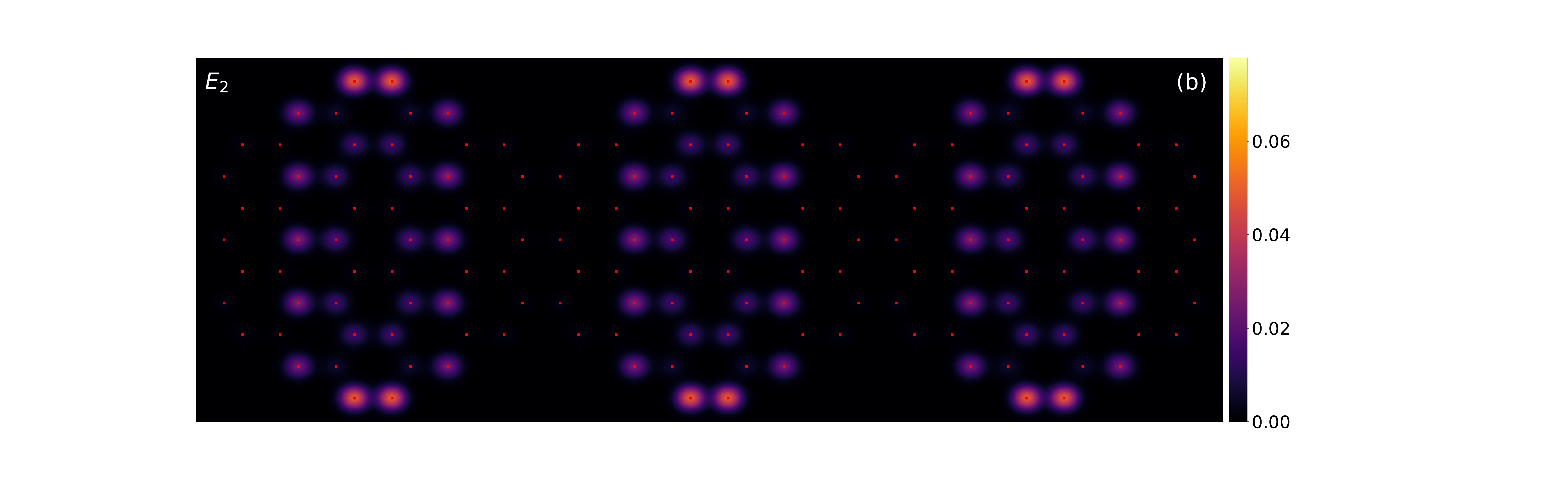}
 \end{minipage}
 \begin{minipage}{.5\textwidth}
  \includegraphics[trim=7cm 2.0cm 8cm 3.1cm, clip=true, width=\textwidth]{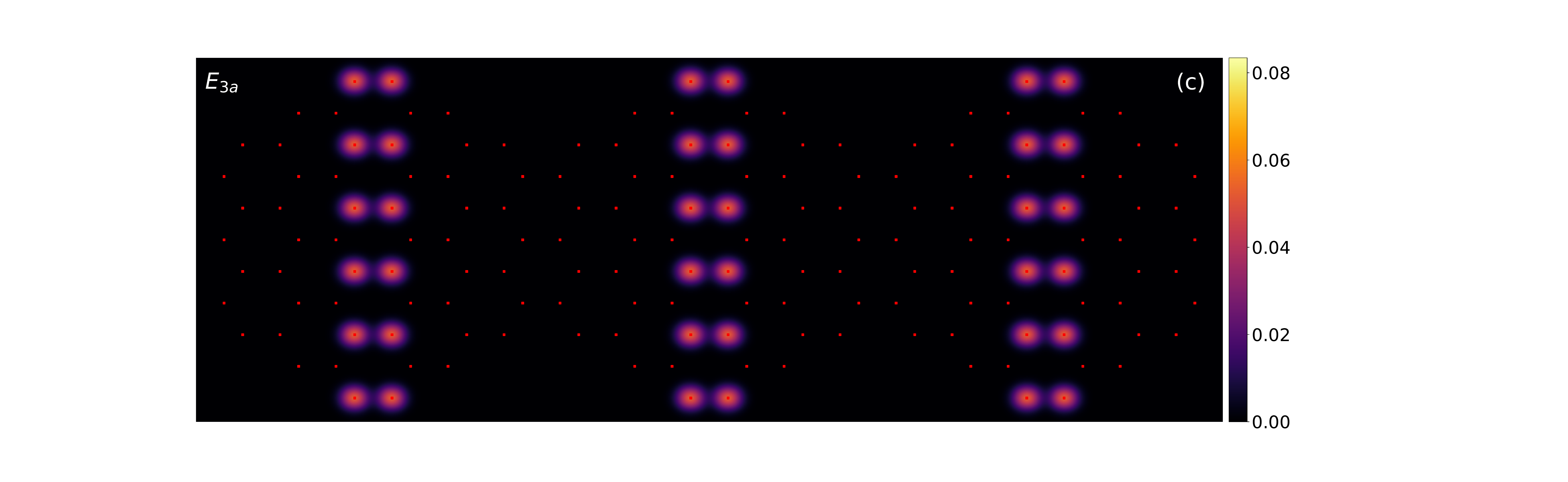}
 \end{minipage}
\caption{Wannier-like states for three flat-bands 
in a 7-AGNR(1,3). 
	}\label{fig7}
\end{figure}

Our tight-binding results for the 3-AGNR(1,3) heterostructure (Fig.~\ref{fig4}) show that these 
Wannier-like $-t$ states, which exist in the odd-N pristine AGNRs, 
survive (basically unaffected) the $(n,m)$ extensions that give origin to the 
heterostructure. This can be seen in the LDOS (charge density) profile shown 
in Fig.~\ref{fig4}(c) for state $E_{3a}$, which shows exactly the same 
structure as the one in Fig.~\ref{fig3}, with the difference that now the 
extended unit cell is wider, thus it accommodates four 
occupied dimers along the vertical direction, in contrast to the pristine 3-AGNR, 
where the Wannier-like state is composed of just two dimers (see Fig.~\ref{fig3}). 
On the other hand, Fig.~\ref{fig4}(d) shows the other Wannier-like state, $E_{3b}$, that 
is degenerate at $E =-t$. Interestingly, its charge profile near the edge of the extended 
unit cell is clearly reminiscent of the pristine 3-AGNR, while, at the center 
of the unit cell it is a mixture of the $E_{3a}$ state and some charge density 
occupying the maximally-separated sites that are left empty by the $E_{3a}$ 
state. 

The interesting result shown in the other panels 
of Fig.~\ref{fig4}, for the remaining three flat-bands [panels (a), (b), and (e)], is 
that they seem to also originate from Wannier-like states with 
different charge configurations [when compared to panels (c) and (d)] that also 
do not have overlap between adjacent unit cells. 
Thus, in principle, they may produce similar magnetic ground states 
as the one theoretically predicted for the $\pm t$ bands in pristine N-AGNRs~\cite{Lin2009}, 
as long as these flat-bands are crossed by dispersive bands. 
Section \ref{sec:DFT} presents a DFT analysis of this possibility. 

\begin{figure*}[ht]
  \includegraphics[width=2.0\columnwidth]{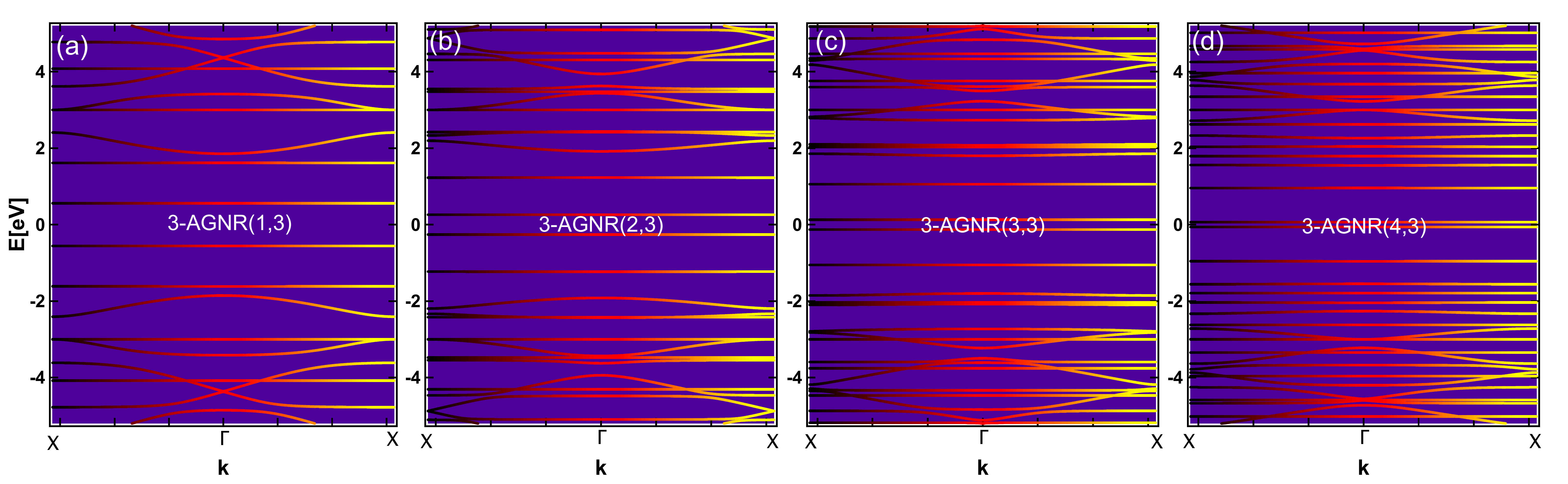}
\caption{Band structures for 3-AGNR(n,3) heterostructures for 
$n=1$ to $4$, in panels (a) to (d), respectively. 
}
\label{fig8}
\end{figure*}

\section{Wannier-like states for $N \geq 5$}\label{sec:vary-N}

In panels (a), (b), and (c) in Fig.~\ref{fig5}, we see the band structure for 
N-AGNR(1,3) heterostructures, for $N=5$, $7$, and $9$, respectively. Despite 
the fact that the complexity of the band structures increases with $N$, we 
can ascertain some facts~\cite{note0}: (i) flat-band $1$, seen in Fig.~\ref{fig2}, 
remains \emph{perfectly} flat for $N=5$ and $7$, although at a different energy position, 
while flat-band $2$ has acquired a tiny dispersion; 
(ii) for all three values of $N$ the $-t$ flat-band is present. 
In reality, as far as we can tell, the $\pm t$ flat-bands occur for any odd value of 
$N$; (iii) for $N=5$ and $7$, flat-band 4 has already acquired some dispersion; 
(iv) likewise, for $N \geq 9$, except for $-t$ flat-band, the other three flat-bands ($1$, $2$ and $4$) 
have acquired dispersion; (v) finally, the $-t$ flat-band for $N=5$ is still double-degenerate, 
while it is not anymore for $N=7$. 
It is possible that farther from the Fermi level ($E=0$, at half-filling) 
there are additional flat-bands (besides the $\pm t$ ones) for $N \geq 9$, 
but we have not investigated this possibility. 

In Figs.~\ref{fig6} and \ref{fig7}, we show the flat-band Wannier-like states  
corresponding to bands $1$, $2$ and $3$ presented in Figs.~\ref{fig5}(a) 
and \ref{fig5}(b), for a 5-AGNR(1,3) and a 7-AGNR(1,3) heterostructure, respectively. 
A careful comparison of Figs.~\ref{fig4}, ~\ref{fig6}, and ~\ref{fig7} 
shows that the Wannier-like states for the same band at different values 
of $N$ are semi-quantitatively the same, indicating that the maximum  
$N$ for which we can look for these interesting states is $N=7$, 
which is an N-AGNR(n,m) heterostructure size that can be faithfully obtained 
in the laboratory~\cite{Rizzo2018,Cai2010,Houtsma2021}, 
suggesting that the results obtained here can be tested experimentally. 

As mentioned above, there is an interesting point regarding 
the $-t$ flat-band Wannier-like states $E_{3a}$ and $E_{3b}$ 
as we vary $N$ in an N-AGNR(1,3) heterostructure: they are still degenerate for 
$N=5$, as can be seen in Fig.~\ref{fig6}(c), where we show the combined charge density 
for both bands $E_{3a}$ and $E_{3b}$, however, for $N=7$, it is not degenerate anymore. 
Notice that in Fig.~\ref{fig7} we show, in panel (c), the charge density just for 
the $E_{3a}$ band, since band $E_{3b}$ does not exist anymore. We speculate that, 
as may be inferred from the charge density distribution in Fig.~\ref{fig6}(c), 
the Wannier-like state $E_{3b}$ for $N=5$ seems on the verge of losing its 
Wannier-like character. This occurs because there are only 4 sites (indicated by numbers 1 to 4, and   
showing perfect destructive quantum interference) preventing the existence of a continuous nearest-neighbor path 
that connects all unit cells with each other, which would result in a dispersive state. 

\begin{figure*}[ht]
  \includegraphics[width=2.0\columnwidth]{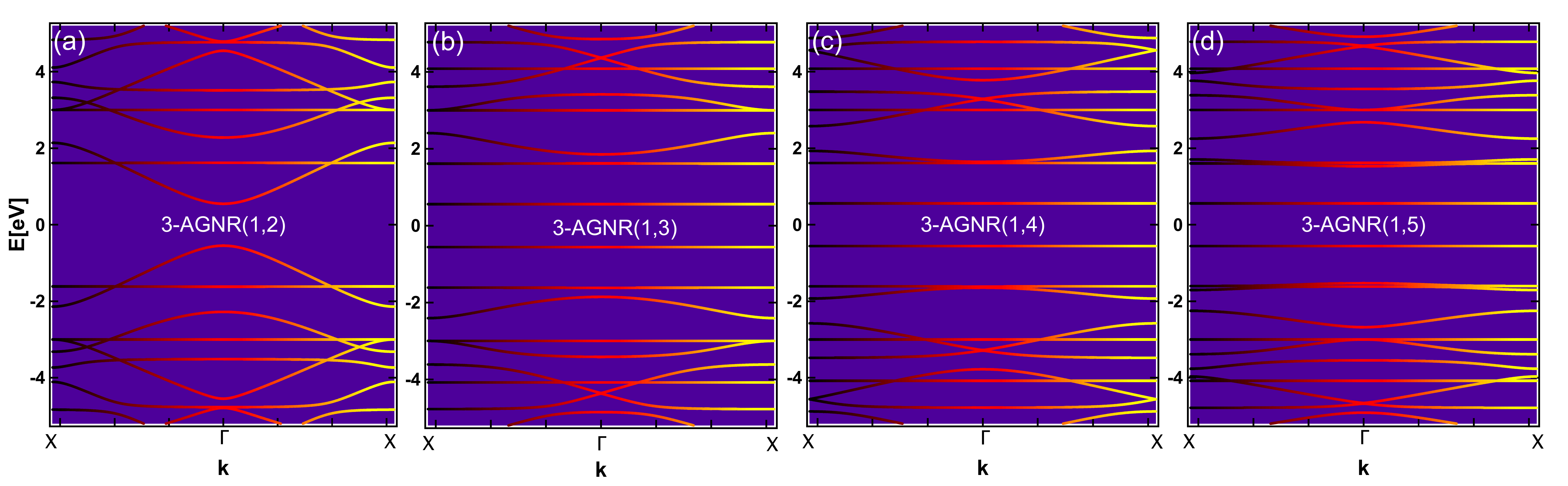}
\caption{Band structures for 3-AGNR(1,m) heterostructures for 
$m=2$ to $5$, in panels (a) to (d), respectively. 
}
\label{fig9}
\end{figure*}

\section{Dependence on parameters $n$ and $m$.}\label{sec:n-and-m}

\subsection{Band structure dependence with $n$}

In Fig.~\ref{fig8}, we see tight-binding band-structure results for 
3-AGNR(n,3), for $n=1$ to $4$, in panels (a) to (d), respectively. 
In panel (a), we repeat the results shown in Fig.~\ref{fig2} [for 3-AGNR(1,3)] 
to facilitate comparison. A trend with increasing $n$ 
(size of the extended region of the heterostructure) can be clearly discerned. 
Indeed, we see that the $\pm t$ flat-bands survive the increase in the unit cell, and 
a cluster of flat-bands (and some bands with very little dispersion) 
develops in the energy range $-2.0 \lesssim E \lesssim -1.0$. It is also interesting to 
remark that flat-band $E_1$ (the one closest to the Fermi energy) tends to approach 
the Fermi energy as $n$ increases. We also did an analysis 
for larger values of $n$. For example, for $n=10$ (not shown), 
bands at higher energies seem to become less dispersive. 
In addition, the flat-band closest to the Fermi energy remains flat and 
approaches the Fermi energy even more, sitting basically 
at the Fermi energy for a 3-AGNR(10,3) heterostructure. Finally, for $n=10$, the 
cluster of flat-bands mentioned above becomes more dense and somewhat 
closer to the Fermi energy. 

We also investigated the band structure dependence with $n$ for 5-AGNR(n,3) 
heterostructures (not shown) and obtained qualitatively the same results 
as the ones shown in Fig.~\ref{fig8} for $N=3$, which may be considered 
reasonable, since we can intuitively expect a lesser dependence of the 
electronic structure on $N$ than on $n$ and $m$. 

\subsection{Band structure dependence with $m$}

In Fig.~\ref{fig9}, we see the band structures for 
3-AGNR(1,m) heterostructures for $m=2$ to $5$ in panels (a) to (d), respectively. 
Here, we also reproduced Fig.~\ref{fig2}, in panel (b), to 
facilitate comparison. As seen with the variation of $n$ (but to a 
lesser degree), we see in Fig.~\ref{fig9}, for 
3-AGNR(1,m), that increasing $m$ from $2$ to $5$ results in an 
accumulation of flat-bands close to the Fermi energy. 
In addition, as observed for the $n$-variation, the results 
for the $m$-variation of the 5-AGNR(1,m) heterostructures (not shown) are 
qualitatively similar to the trend seen in Fig.~\ref{fig9} for 3-AGNR(1,m). 

We wish to call attention to the band structure 
in Fig.~\ref{fig9}(a), for 3-AGNR(1,2). In it, we see that the flat-band 
closest to the Fermi energy is crossed by a dispersive band that may 
be topologically non-trivial~\cite{Groning2018,Rizzo2018}. In case 
this dispersive band is indeed topologically non-trivial, it would be very 
interesting to study the interplay of topology and ferromagnetism once 
this system is doped. 

Before presenting the DFT results, we compile below the results presented in 
Figs.~\ref{fig4} to \ref{fig9}. This may serve as a guide to the reader 
to relate the presence (or absence) and behavior of flat-bands 
with the variation of parameters $N$, $n$ and $m$:

\begin{enumerate}

\item The $\pm t$ flat-bands, present in the pristine AGNRs, are also 
present for \emph{all} values of $N$, $n$, and $m$ investigated 
here, and they are associated to the \emph{same} Wannier-like 
states identified in the pristine AGNRs~\cite{Lin2009}.

\item For N-AGNR(1,3) ($N=3$ and $5$), the $\pm t$ bands 
are double degenerate (in contrast to the pristine AGNRs) and the 
partner state is also a Wannier-like state, similar to the one 
mentioned in the item above. This degeneracy is lifted for $N>5$. 

\item Additional flat-bands appear around the $\pm t$ flat-bands 
for all heterostructures analyzed, and to each different flat-band 
it was possible to associate a Wannier-like state that seems like 
a variant of the $\pm t$ Wannier-like state. 

\item Regarding the variation of these additional flat-bands with $N$, 
we see that they survive (i.e., have zero-dispersion) up to $N=7$ 
for all heterostructures studied here. 

\item With increasing $n$, we see that the overall number 
of flat-bands increases, with a cluster of them forming 
gradually closer to the Fermi-energy, with one of them 
seating almost \emph{at} the Fermi energy already for the 3-AGNR(10,3) 
heterostructure. This description of the $n$ dependence applies to 
all prime values $3 \leqslant N \leqslant 7$. 

\item Similar to the $n$-dependence, there is an increase in the 
number of flat-bands with $m$, with a similar accumulation close to 
the Fermi energy. As well, this description qualitatively applies to 
all prime values $3 \leqslant N \leqslant 7$.

\end{enumerate}

We should also mention that a brief study of the so-called `Staggered' 
heterostructures, which are less symmetric than the ones analyzed here 
(see Refs.~\cite{Groning2018,Rizzo2018}), has shown a tendency to form 
considerably less flat-bands, indicating that the hetrostructures discussed 
here are the ones that should receive more attention in the quest for 
quasi-1D ferromagnetism. 

\section{Ferromagnetic phase obtained with DFT}\label{sec:DFT}

To address the possible existence of any magnetic phase under hole-doping, 
we will use DFT, which is a more realistic calculation than tight-binding 
and that can treat correlations at the mean-field level. 
We will search for indications of a ferromagnetic 
ground-state on two heterostructures, viz., 3-AGNR(1,3) and 5-AGNR(1,3).
According to Ref.~\cite{Lin2009}, the presence of itinerant carriers is 
important to mediate ferromagnetism between the isolated magnetic moments 
in each unit cell of the Wannier-like states. The 3-AGNR(1,3) and 5-AGNR(1,3) 
heterostructures present dispersive bands intercepting the flat-bands, 
as can be seen in Figs.~\ref{fig2} and \ref{fig5}(a), respectively.  
We will postpone a careful DFT analysis of the ferromagnetic 
ground-state dependence on the parameters $n$ and $m$ to a future publication.

\subsection{Details of the DFT calculations}

We do a DFT calculation within the projector  augmented  wave scheme~\cite{Blochl1994} 
for the pseudopotentials. The total energies and
electronic structures are self-consistently computed within  a plane-wave
basis-set with a kinetic  energy cut-off  of 350~eV.  We  used the
Vienna Ab initio Simulation Package (VASP)~\cite{Kresse1996,PRB.54.11169}.
For a better description of the exchange-correlation term of the DFT,
we use a hybrid functional to improve the description of the many-electron
interactions and charge localization~\cite{Heyd2003}.
The HSE06 hybrid functional has been used~\cite{Krukau2006},
where the screened functional contains part of the exact Hartree-Fock exchange
that has been shown to give accurate results for the exchange splitting, which is 
crucial to understand the magnetic properties in our system.
Interestingly, our results show that the inclusion 
of the hybrid functional puts the $E_{3a}$ flat-band
around 3~eV from the Fermi energy, matching the tight-binding results (see Fig.~\ref{fig10}). 
By suppressing the hybrid functional, using just the generalized gradient 
approximation~\cite{PRL.77.3865}, the $E_{3a}$ band stays around 2.5~eV from the Fermi level.
As we are using the periodic supercell approach within the first principles calculations,
the exchange interactions between adjacent unit cells are also included.

\subsection{Band structure for 3-AGNR(1,3): comparison DFT/tight-binding} 

\begin{figure}
\includegraphics[trim=0cm 0cm 0cm 0cm,clip=true,width=0.45\textwidth,angle=0.]{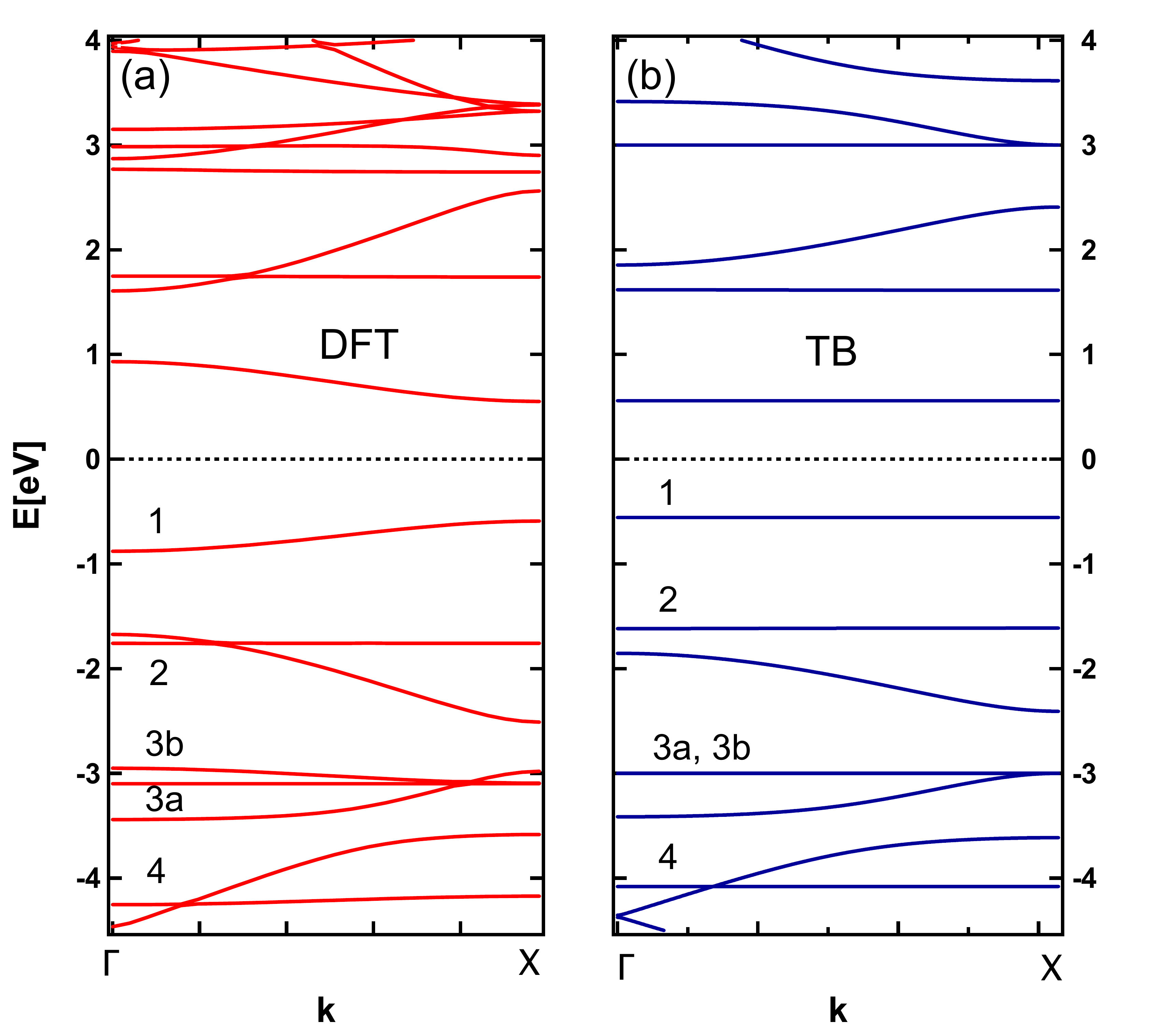}
	\caption{DFT and tight-binding band structures for a 3-AGNR(1,3) heterostructure, 
	at half-filling, in panels (a) and (b), respectively. 
	As expected, the DFT bands are not particle-hole symmetric, 
	but, other than that, there is a good qualitative agreement between DFT and 
	tight-binding. The numbered bands are discussed in the text. 
}
\label{fig10}
\end{figure}

Panels (a) and (b) in Fig.~\ref{fig10} show a comparison of the DFT and tight-binding 
band structures for a 3-AGNR(1,3) heterostructure, respectively. Contrary to the tight-binding 
bands, the DFT bands are not mirror symmetric around $E=0$. Note that the tight-binding bands 
would also lack mirror symmetry if a next-nearest-neighbor hopping had been introduced 
(breaking chiral symmetry). Some details of the negative energy DFT bands are worthy 
of mention. First, we see that the DFT band closest to the Fermi energy 
(numbered $1$ in Fig.~\ref{fig10}), which is flat 
in the tight-binding results, has acquired dispersion. Fig.~\ref{fig4}(a) 
shows the tight-binding Wannier-like state for this band. Since its charge density is 
mostly accumulated at the edges of the unit cell (and it does not completely vanish at 
its center either), one may argue that small 
perturbations introduced by the DFT calculations to the tight-binding results may create 
an overlap between the Wannier-like states in adjacent unit cells and result in dispersion 
(as discussed above in relation to the $E_{3b}$ tight-binding band for a 5-AGNR(1,3) heterostructure). 
On the other hand, the Wannier-like states [see panels 
(b) and (c) in Fig.~\ref{fig4}] for the bands denoted $2$ and $3a$ in Fig.~\ref{fig10} 
are much more concentrated at the 
center of the unit cell [especially for band $3a$, see Fig.~\ref{fig4}(c)] 
and thus they should be more robust against perturbations that could create an overlap 
between adjacent unit cells. Thus, as expected, DFT bands $2$ and $3a$ are \emph{perfectly} flat. 
Finally, the same reasoning leads us to expect that the DFT bands $3b$ and $4$ should acquire 
dispersion, as they do indeed, the latter less so than the former.  

A final point can be made, along the lines of the qualitative discussion above, 
if we compare our DFT results with the DFT results in Ref.~\cite{Lin2009}. There, it was obtained, 
for a pristine (no extensions) 5-AGNR, that the $\pm t$ DFT flat-bands, at zero doping, 
acquire a dispersion of $\approx 0.4$~eV (see Fig.~4(a) in Ref.~\cite{Lin2009}). 
On the other hand, the DFT $\pm t$ bands for N-AGNR(1,3), for $N=3$ [band $3a$ in Fig.~\ref{fig10}(a)] 
and $N=5$ (not shown), are \emph{perfectly} flat. 
This seems to indicate that in an N-AGNR(n,m) heterostructure, which has a wider unit 
cell than a pristine AGNR, the charge density of the $\pm t$ Wannier-like states in each unit cell 
[like the ones shown in Figs.~\ref{fig4}(c), \ref{fig6}(c), and \ref{fig7}(c)] 
is even more insulated from the charge density in adjacent unit cells, and thus can result 
in a more robust (more massive) DFT flat-band. 

\subsection{DFT bands at finite doping and ferromagnetic ground-state}

To bring the Fermi energy close to the flat-bands, and thus investigate their properties, 
we start hole-doping the 3-AGNR(1,3) 
heterostructure. We measure the hole-doping $x_h$ from the half-filling point, thus 
$x_h=1-\langle n \rangle$ (therefore, $x_h=0$ at half-filling), 
where $\langle n \rangle$ is the electron average site-occupancy. 

\begin{figure}[ht]
  \includegraphics[trim=0cm 0cm 0cm 0cm,clip=true,width=0.5\textwidth]{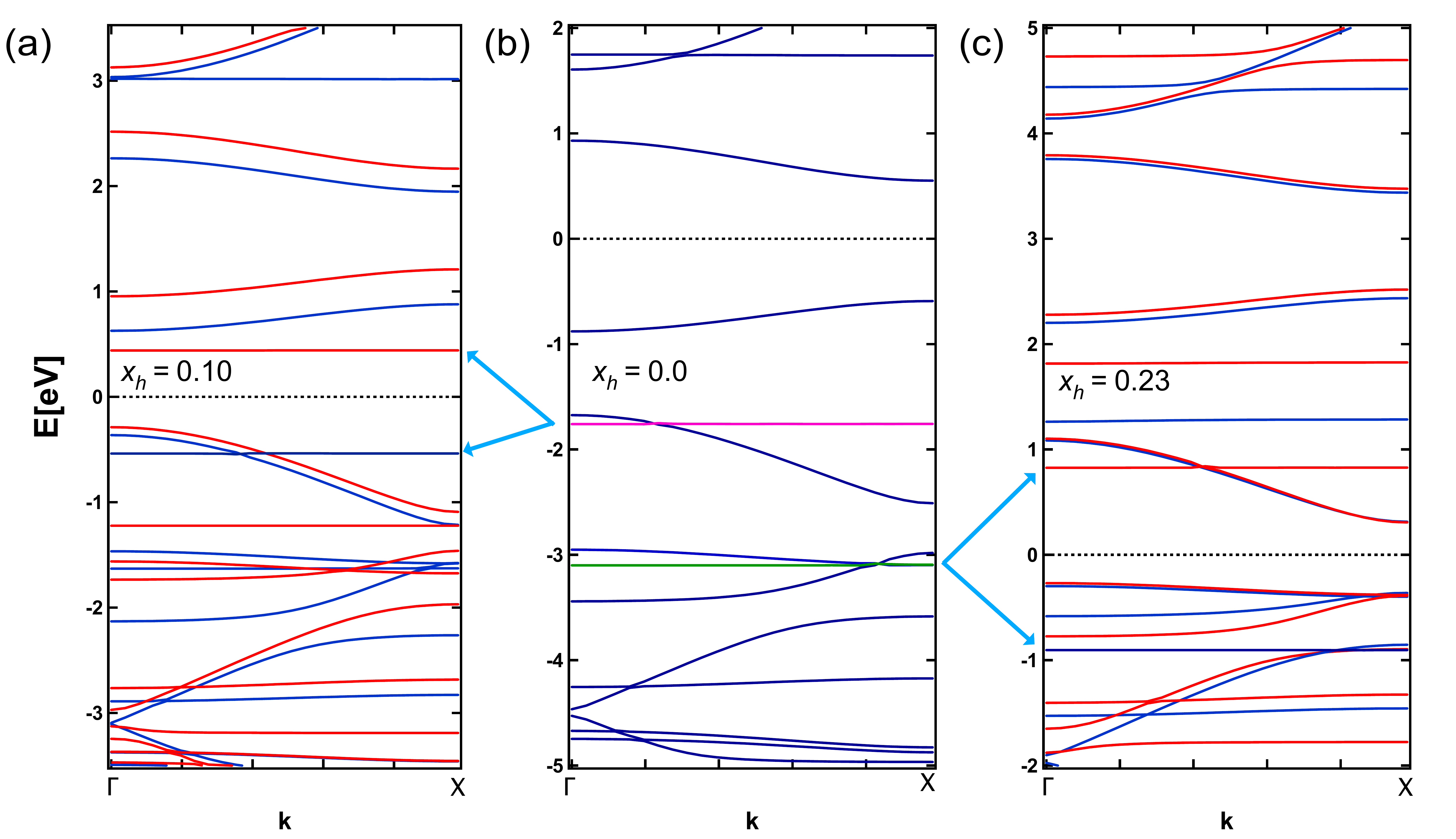}
\caption{DFT band structures for a 3-AGNR(1,3) heterostructure at different 
	hole-dopings: (a) $x_h=0.10$, (b) $x_h=0.0$ (half-filling), and 
	(c) $x_h=0.23$. In panels (a) and (c), majority-spin bands are in blue and 
	minority-spin bands are in red. 
}
\label{fig11}
\end{figure}

In Fig.~\ref{fig11}, we show the DFT bands for $x_h=0.10$, $0.0$ (half-filling), and 
$0.23$, in panels (a) to (c), respectively. The Fermi energy is at $E=0.0$ in each panel. 
In panel (b), we repeat the results shown in Fig.~\ref{fig10}(a) to better illustrate 
the hole-doping effects. In Figs.~\ref{fig11}(a) and \ref{fig11}(c), at finite doping, 
we show the spin-decomposed band structure obtained through a hybrid DFT calculation,
where the majority-spin bands are denoted in blue and the minority-spin 
bands are in red. The cyan arrows connecting the center panel to each one of the adjacent 
panels indicate the extent of the exchange splitting of each flat-band. The arrows 
connecting band $2$ [in panel (b)] to the corresponding exchange-split bands in 
panel (a) indicate the extent of the exchange splitting energy acting over band $2$ 
for $x_h=0.1$, given by $E_{2,0.1} \approx 1.0$~eV. Likewise, the arrows connecting 
panels (b) and (c) indicate the exchange splitting energy of band $3a$ for $x_h=0.23$, 
corresponding to $E_{3a,0.23} \approx 2.0$~eV. 

\begin{figure}[ht]
  \includegraphics[width=0.8\columnwidth]{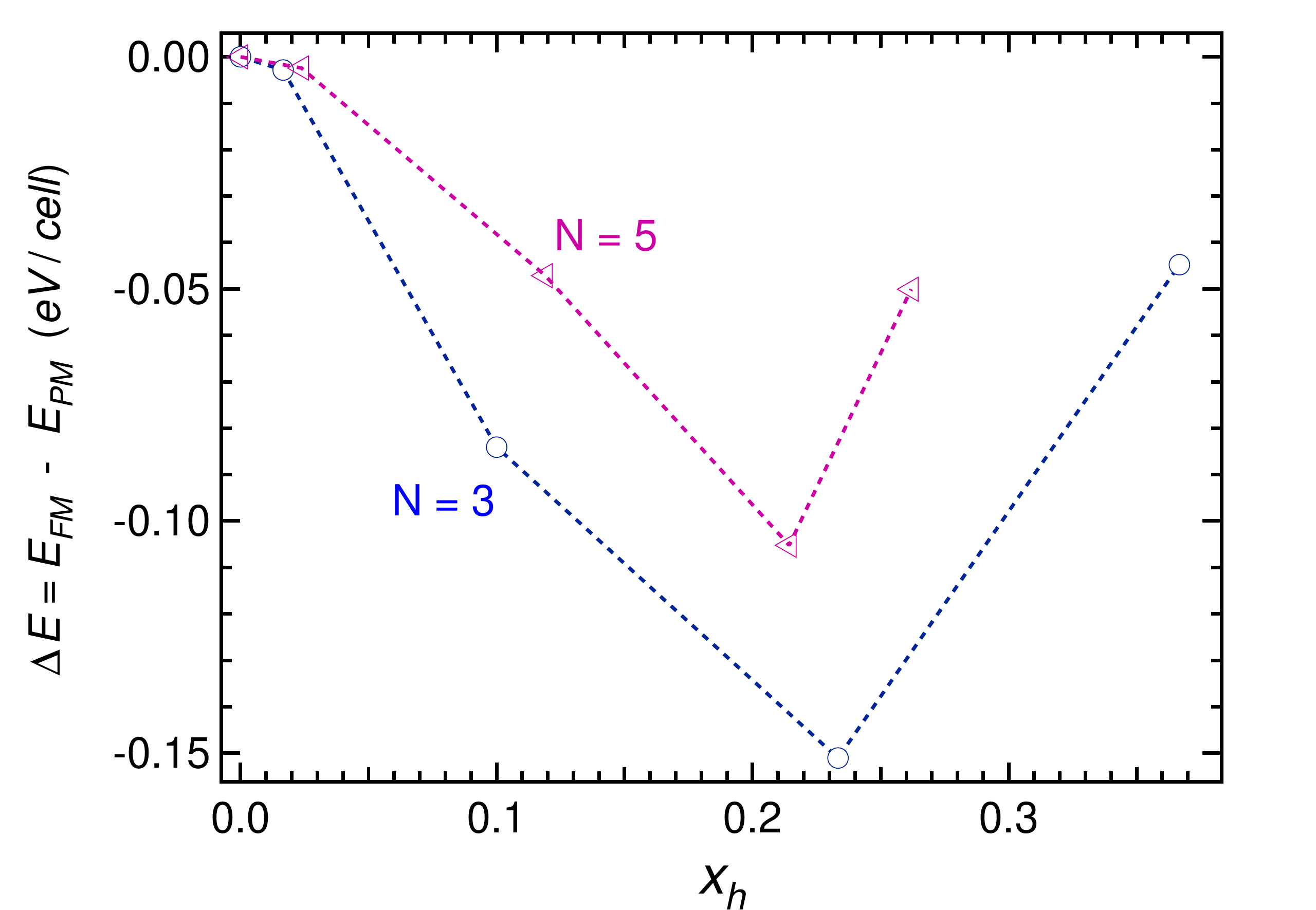}
	\caption{$\Delta E = E_{FM} - E_{PM}$ as a function 
	of hole-doping $x_h$ for 3-AGNR(1,3) (blue circles) and 
	5-AGNR(1,3) (purple left-triangles). 
}
\label{fig12}
\end{figure}

In Fig.~\ref{fig12}, we show the energy difference between the ferromagnetic and 
paramagnetic states,  $\Delta E = E_{FM} - E_{PM}$, for both a 3-AGNR(1,3) (blue circles) and a 
5-AGNR(1,3) (purple left-triangles), where $\Delta E < 0$ indicates a ferromagnetic ground-state. 
The most stable ferromagnetic configuration occurs when the 
hole-doping reaches the $3a$ flat-band, for both 3- and 5-AGNR(1,3). The inverse dependence
of the ferromagnetic stability with $N$ can be attributed to the reduction of the overall band 
flatness as $N$ increases (see Figs.~\ref{fig2} and \ref{fig5}). 

From Ref.~\cite{Lin2009}, we obtain that the gain in energy due to ferromagnetic 
ordering of a pristine 5-AGNR is $\Delta E_p \approx -37.5$~meV (per unit cell). 
Since the number of occupied Carbon atoms in the ferromagnetic state in each unit cell is 
$N_{occ} = 6$ (see Fig.~5(b) in Ref.~\cite{Lin2009}), we obtain 
$\nicefrac{\Delta E_p}{N_{occ}} = -6.25$~meV. 
The corresponding results for the two heterostructures we analyzed through 
DFT, i.e., 3-AGNR(1,3) and 5-AGNR(1,3), were $\Delta E_3=-150$~meV, 
$N_{occ}=8$, and $\Delta E_5=-105$~meV, $N_{occ}=10$. This results 
in $\nicefrac{\Delta E_N}{N_{occ}}=-18.8$~meV and $-10.5$~meV, respectively. 
This shows that, if we compare the ferromagnetic energy gain for the pristine 
5-AGNR and the 5-AGNR(1,3), the heterostructure had almost 70\% 
more energy gain than that of the pristine AGNR. 
We believe that to be the case for two main reasons. First, the N-AGNR(n,m) heterostructures 
studied here through DFT present true flat-bands, contrary to what was seen in the pristine N-AGNRs 
studied in Ref.~\cite{Lin2009}. Second, the pristine N-AGNRs show a single \emph{low-dispersion} band, 
while our N-AGNR(n,m) heterostructures show multiple \emph{perfectly} flat-bands [two in the 
case of 3-AGNR(1,3), bands labeled 2 and $3a$ in Fig.~\ref{fig10}(a)] and multiple \emph{almost} 
flat-bands [two in the case of 3-AGNR(1,3), bands $3b$ and 4 in Fig.~\ref{fig10}(a)], which should 
clearly result in a more robust ferromagnetic ground state. 

\section{Wannier-like states: comparison between DFT and tight-binding}

In this section, we want to highlight the fact that it is not only the 
DFT and tight-binding band structures that are qualitatively similar 
(as shown in Fig.~\ref{fig10}), but also the Wannier-like states associated 
with the flat-bands obtained by either method that are qualitatively 
similar too. 

\begin{figure}
 \begin{minipage}{.5\textwidth}
\includegraphics[trim=7cm 1.2cm 6.0cm 1.3cm, clip=true, width=\textwidth]{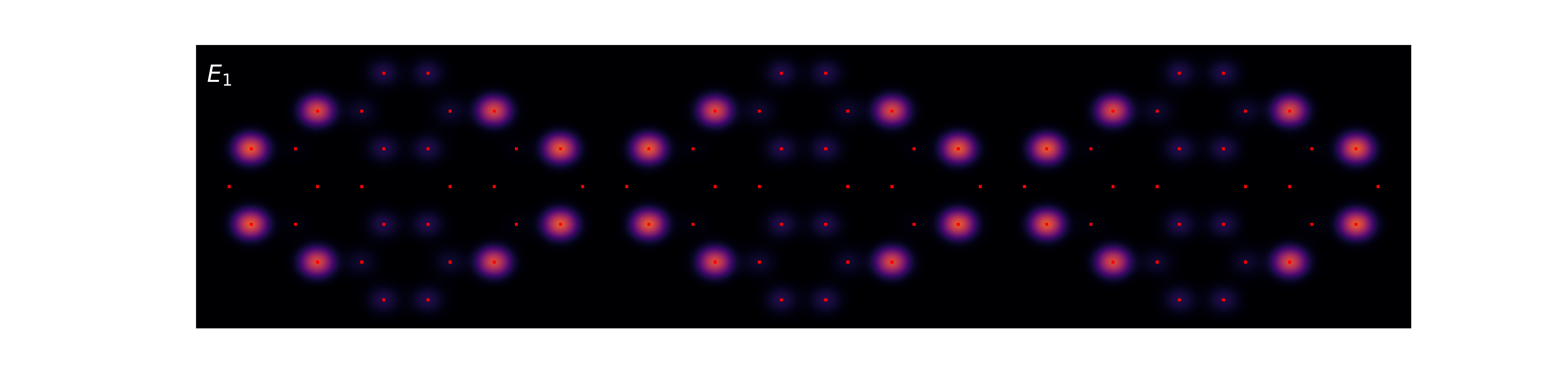}
 \end{minipage}
 \begin{minipage}{.5\textwidth}
\includegraphics[width=0.98\textwidth]{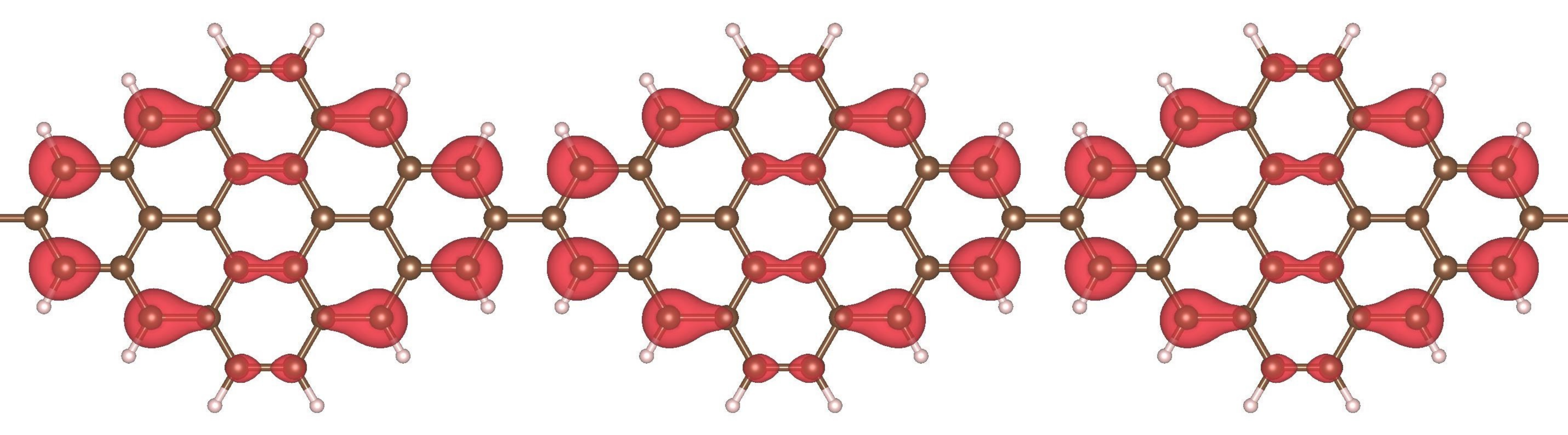}
 \end{minipage}
\caption{Wannier-like state for flat-band $E_1$ for a 3-AGNR(1,3) 
	heterostructure. 
	Top panel: tight-binding result; bottom panel: DFT result. 
}\label{fig13}
\end{figure}

\begin{figure}
 \begin{minipage}{.5\textwidth}
\includegraphics[trim=7cm 1.2cm 6cm 1.3cm, clip=true, width=\textwidth]{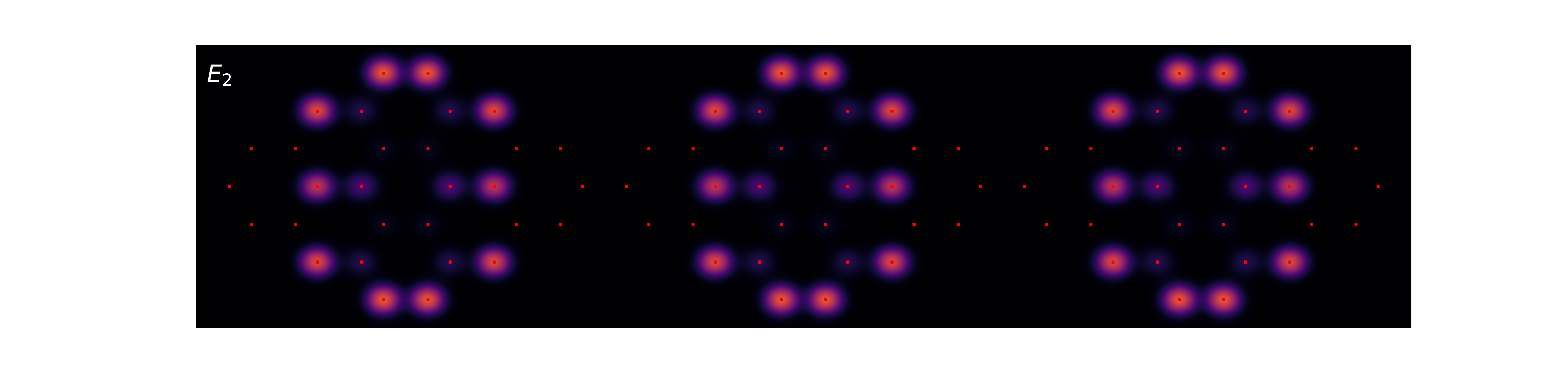}
 \end{minipage}
 \begin{minipage}{.5\textwidth}
\includegraphics[width=0.98\textwidth]{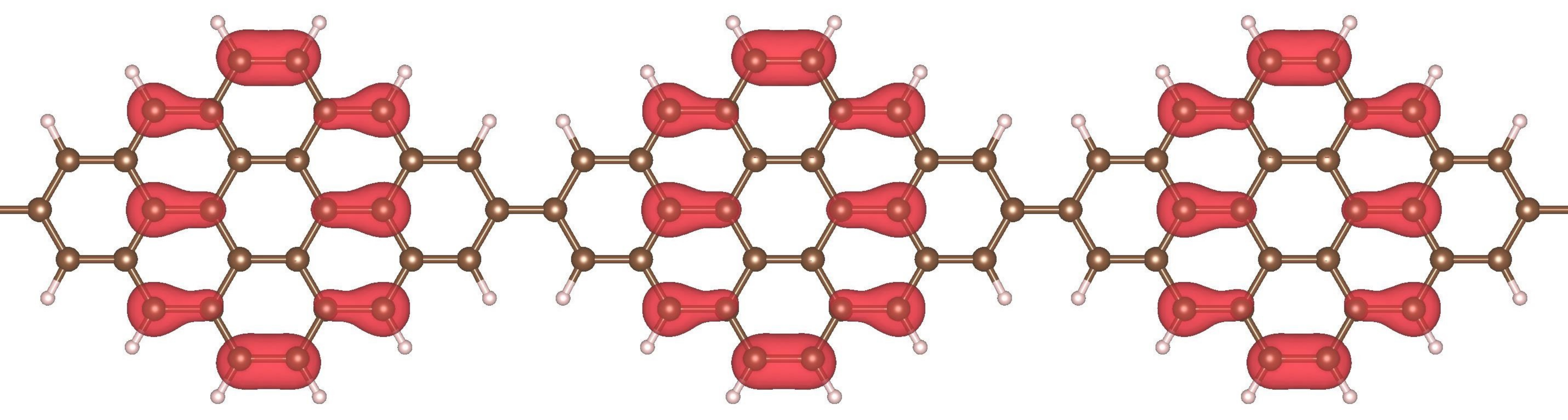}
 \end{minipage}
 \begin{minipage}{.5\textwidth}
\includegraphics[trim=7cm 1.2cm 6cm 1.3cm, clip=true, width=\textwidth]{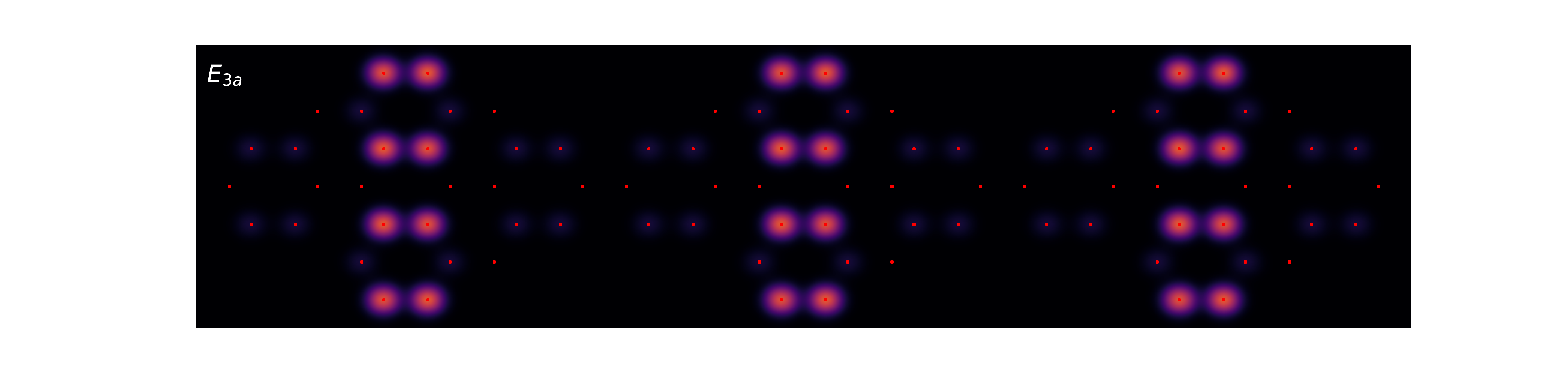}
 \end{minipage}
 \begin{minipage}{.5\textwidth}
\includegraphics[width=0.98\textwidth]{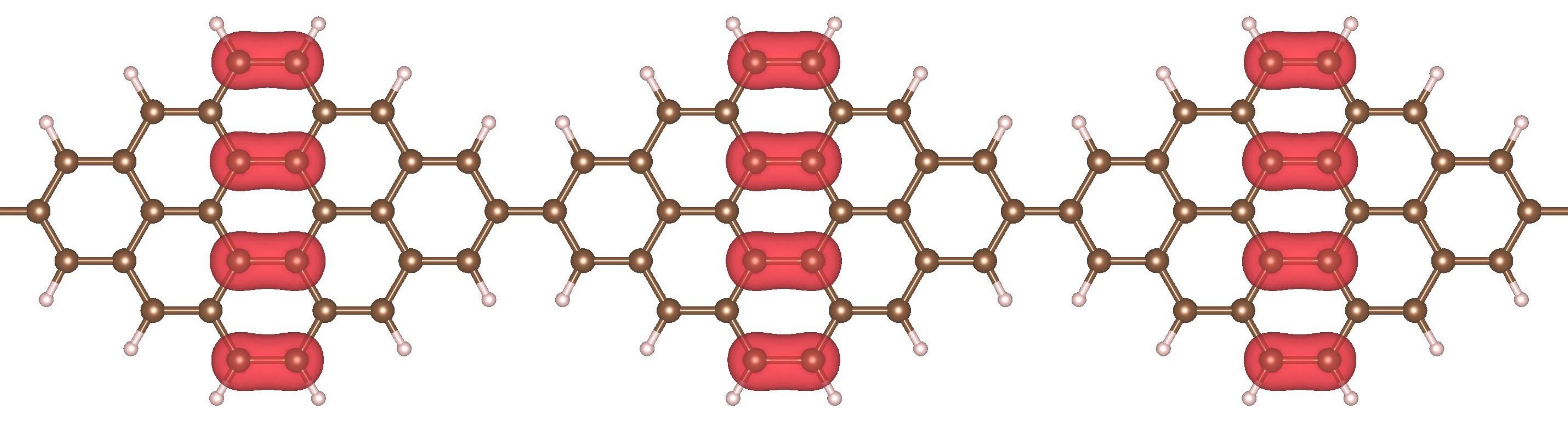}
 \end{minipage}
 \begin{minipage}{.5\textwidth}
\includegraphics[trim=7cm 1.2cm 6cm 1.3cm, clip=true, width=\textwidth]{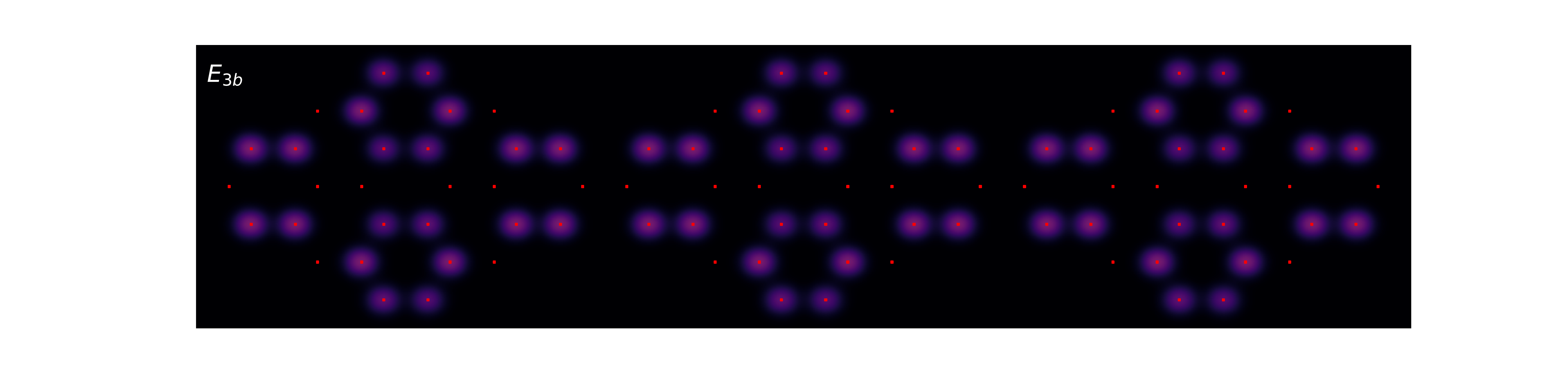}
 \end{minipage}
 \begin{minipage}{.5\textwidth}
\includegraphics[width=0.98\textwidth]{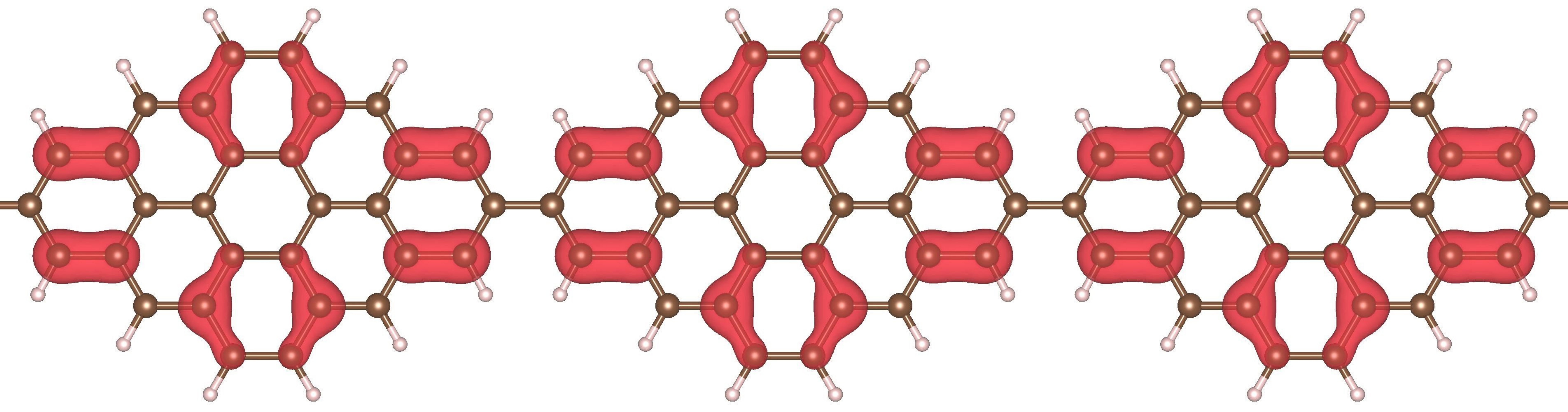}
 \end{minipage}
 \begin{minipage}{.5\textwidth}
\includegraphics[trim=7cm 1.2cm 6cm 1.3cm, clip=true, width=\textwidth]{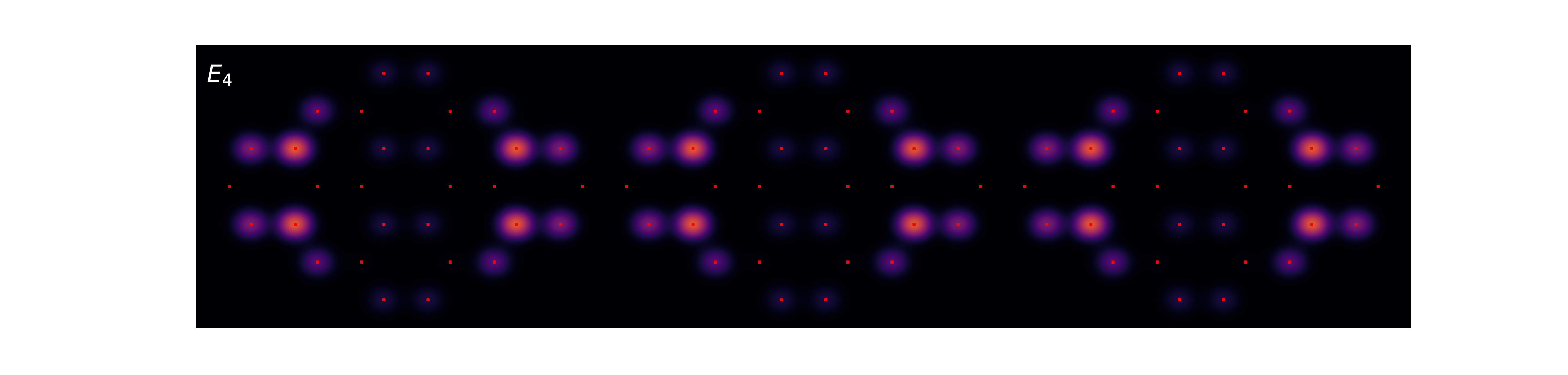}
 \end{minipage}
 \begin{minipage}{.5\textwidth}
\includegraphics[width=0.98\textwidth]{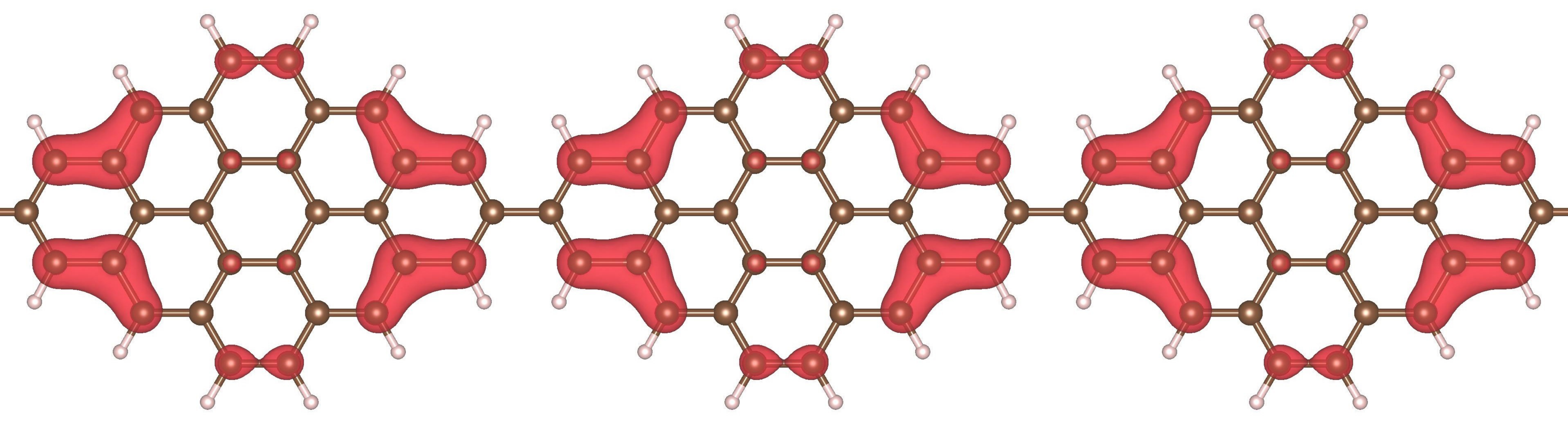}
 \end{minipage}
	\caption{Wannier-like states for flat-bands $E_2$, $E_{3a}$, 
	$E_{3b}$, and $E_4$ for a 3-AGNR(1,3) heterostructure. 
	Top panel: tight-binding result; bottom panel: DFT result, for all 
	pairs of results. 
}\label{fig14}
\end{figure}

In the top panel of Fig.~\ref{fig13} we reproduce Fig.~\ref{fig4}(a), with the 
tight-binding result for the flat-band $E_1$ Wannier-like state for a 3-AGNR(1,3) at half-filling. 
In the bottom panel, we show the corresponding DFT result. Close inspection 
indicates that there is a semi-quantitative agreement between tight-binding and DFT. 
Figure~\ref{fig14} makes the same comparison for flat-bands 
$2$, $3a$, $3b$, and $4$, and close inspection of the plots shows that the tight-binding 
results are surprisingly close to the DFT results in all cases. 

\section{Summary and conclusions}\label{sec:sum$conc}
We have used the tight-binding and DFT methods to study the electronic 
properties of recently synthesized N-AGNR(n,m) graphene heterostructures~\cite{Groning2018,Rizzo2018}, 
which have been shown to present, for specific values of $N$, $n$, and $m$, 
topological properties at low energy that can be simulated by the SSH model. 
We found out that the heterostructures show a multiplicity of flat-bands, 
whose properties can be reasonably well controlled by the parameters 
$N$, $n$, and $m$. We see flat-bands in our heterostructures up to 
$N = 7$. We have strong indications that the quantum interference mechanism 
that gives origin to the $\pm t$ single flat-band in pristine AGNRs~\cite{Lin2009} is at 
play in all the flat-bands analyzed in our heterostructures. 
The pristine AGNR $\pm t$ bands are still present in the heterostructures, but 
with the interesting presence of a degenerate partner  (for $N = 3$ and $5$) 
in the tight-binding simulations. This degeneracy is slightly lifted in the 
DFT results for all values of $N$. Importantly, our DFT results show that 
a few of the flat-bands observed in the tight-binding simulations remain 
\emph{perfectly} flat in the DFT simulations as well. Thus, the ferromagnetism 
observed in our DFT results is considerably stronger than 
that observed in pristine AGNRs~\cite{Lin2009}. 
As a bonus, we found that the charge density associated with the flat-bands 
obtained via tight-binding agree surprisingly well with the corresponding 
results obtained through DFT.

Given the experimental availability of these heterostructures, our results 
suggest that it would be interesting to experimentally explore the possibility 
of ferromagnetism in these systems, which, given the variety of parameters 
that can be manipulated, opens up the possibility of looking for  
non-trivial topology in a ferromagnetic quasi-1D system. 

\section{Acknowledgments.}

P.A.A. thanks the Brazilian funding agency CAPES 
for financial support and L.S.S. acknowledges support from PROGRAD/UFU.
T.M.S. acknowledges INCT in Carbon Nanomaterials, CNPq, FAPEMIG,
and the computational facilities from LNCC and Cenapad.
G.B.M. acknowledges financial support from the Brazilian agency Conselho Nacional de 
Desenvolvimento Cient\'{\i}fico e Tecnol\'ogico (CNPq), processes 424711/2018-4, 305150/2017-0 and 210 355/2018.

\appendix

\section{Tight-binding Hamiltonian for an N-AGNR(1,3) heterostructure}\label{ap:seca1}
\begin{figure}[ht]
\includegraphics[width=0.5\textwidth]{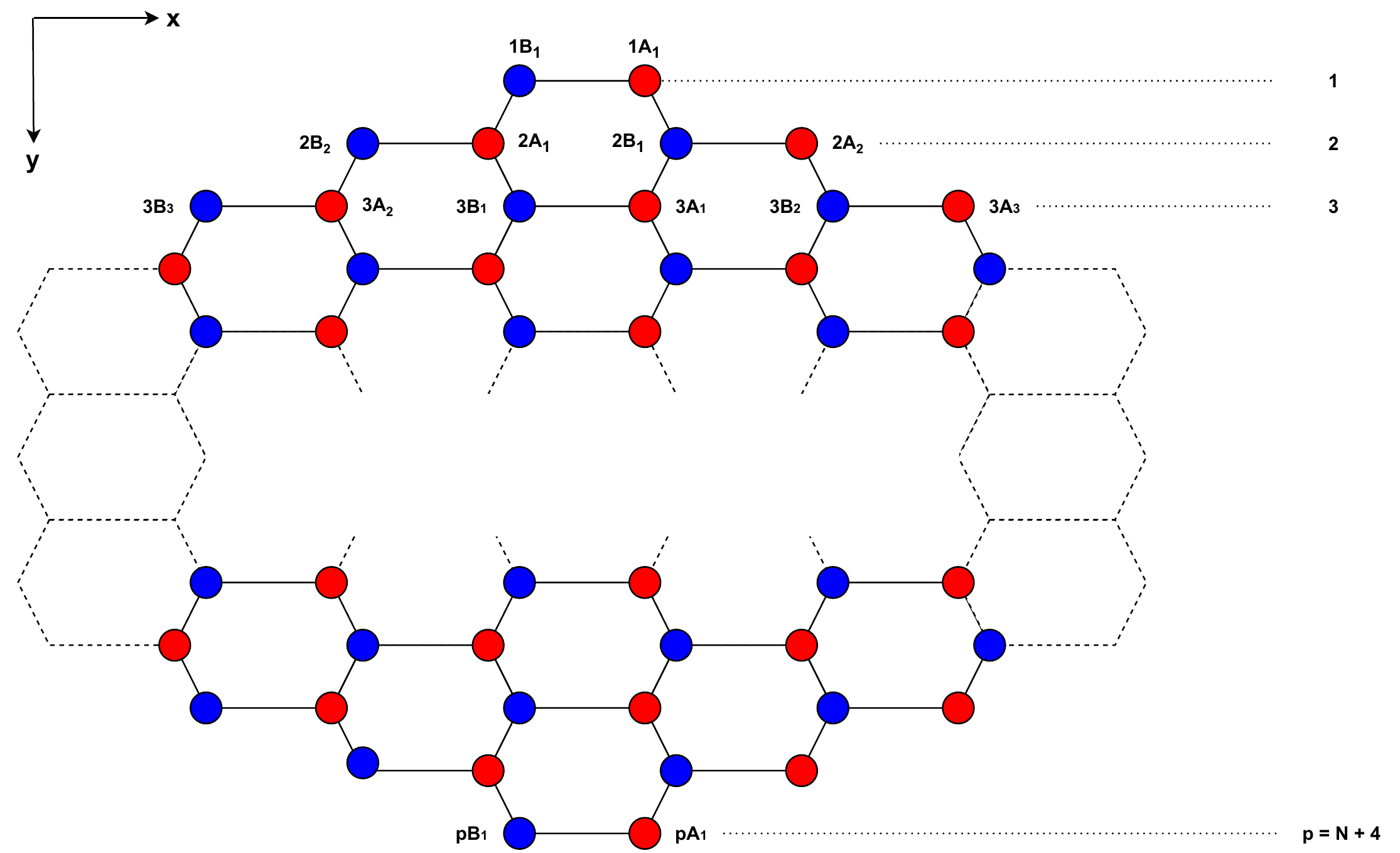}
\caption{Site labelling of a unit cell for an N-AGNR(1,3) heterostructure. 
The sites are labeled $p\alpha_q$, where $\alpha=\rm{A/B}$, $1 \leq p \leq N+4$  
and $1 \leq q \leq 3$, see text for details. 
}
\label{figA1}
\end{figure}

In this Appendix, we present explicit expressions for the Hamiltonian of an N-AGNR(1,3) heterostructure, in 
the real and reciprocal spaces. The modifications necessary 
to obtain the Hamiltonian for a general N-AGNR(n,m) heterostructure are straightforward. 
In Fig.~\ref{figA1}, we show the $l$-th unit cell of an N-AGNR(1,3) heterostructure, where the A sublattice 
is represented by blue solid dots and the B sublattice by red solid dots. 
The sites are labeled $p\alpha_q$, where $\alpha=\rm{A/B}$, with $1 \leq p \leq N+4$ 
and $1 \leq q \leq 3$, where $p$ runs along the $y$-direction, as indicated in the right-hand side, 
and $q$ runs along the $x$-direction (starting at the center of the unit cell 
and moving to its borders). 

Using the labeling defined above, we can write the N-AGNR(1,3) Hamiltonian in 
real space as 
\begin{widetext}
	\begin{equation}
	\begin{aligned}
	H=&-t \sum_{l}\left[\sum_{  p\in \text {odd}}^{N} b_{l,1}^{\dagger}(p) a_{l,1}(p)+\sum_{p=2 \atop }^{N-1} b_{l,1}^{\dagger}(p+1) a_{l,1}(p)+\sum_{p=2}^{N-1} a_{l,1}^{\dagger}(p+1) b_{l,1}(p)+\mathrm{H.C.}\right] \\
			&-t \sum_{l}\left[\sum_{p=2 \atop }^{N-2} b_{l,2}^{\dagger}(p+1) a_{l,2}(p)+\sum_{p=2}^{N-2} a_{l,2}^{\dagger}(p+1) b_{l,2}(p)+\mathrm{H.C.}\right] \\
					&-t \sum_{l}\left[\sum_{p=2 \atop   m\in \text {even}}^{N-1} b_{l,1}^{\dagger}(p) a_{l,2}(p)+\sum_{ p=2 \atop   m\in \text {even}}^{N-1} b_{l,2}^{\dagger}(p) a_{l,1}(p)+\mathrm{H.C.}\right] \\
								&-t \sum_{l}\left[\sum_{p=3 \atop }^{N-3} b_{l,3}^{\dagger}(p+1) a_{l,3}(p)+\sum_{p=3}^{N-3} a_{l,3}^{\dagger}(p+1) b_{l,3}(p)+\mathrm{H.C.}\right] \\
					&-t \sum_{l}\left[\sum_{ p=3 \atop  m\in \text {odd}}^{N-2} b_{l,2}^{\dagger}(p) a_{l,3}(p+1)+\sum_{  p=3 \atop  m\in \text {odd}}^{N-2} b_{l,3}^{\dagger}(p+1) a_{l,3}(p)+\sum_{  p=4 \atop  m\in \text {even}}^{N-3} b_{l-1,3}^{\dagger}(p) a_{l,3}(p)+\mathrm{H.C.}\right],
	\end{aligned}
	\label{equ1}
	\end{equation}
\end{widetext}

\noindent
where $a_{l,q}(p)$ [$b_{l,q}(p)$] annihilates an electron on site $pA_q$ ($pB_q$) 
on the $l$-th unit cell. Assuming periodic boundary conditions along the $x$-direction, 
we take a Fourier transform along that direction and obtain the reciprocal space Hamiltonian 

\begin{widetext}
	\begin{equation}
	\begin{aligned}
	H=&-t \sum_{k}\left[\sum_{  m\in \text {odd}}^{N} v_1\beta_{k,1}^{\dagger}(p) \alpha_{k,1}(p)+\sum_{p=2 \atop }^{N-1} v_2\beta_{k,1}^{\dagger}(p+1) \alpha_{k,1}(p)+\sum_{p=2}^{N-1} v_3\alpha_{k,1}^{\dagger}(p+1) \beta_{k,1}(p)+\mathrm{H.C.}\right] \\
	&-t \sum_{k}\left[\sum_{p=2 \atop }^{N-2} v_2\beta_{k,2}^{\dagger}(p+1) \alpha_{k,2}(p)+\sum_{p=2}^{N-2} v_3\alpha_{k,2}^{\dagger}(p+1) \beta_{,2}(p)+\mathrm{H.C.}\right] \\
	&-t \sum_{k}\left[\sum_{p=2 \atop   m\in \text {even}}^{N-1} v_1\beta_{k,1}^{\dagger}(p) \alpha_{k,2}(p)+\sum_{ p=2 \atop   m\in \text {even}}^{N-1} v_1\beta_{k,2}^{\dagger}(p) \alpha_{k,1}(p)+\mathrm{H.C.}\right] \\
	&-t \sum_{k}\left[\sum_{p=3 \atop }^{N-3} v_1\beta_{k,3}^{\dagger}(p+1) \alpha_{k,3}(p)+\sum_{p=3}^{N-3} v_1\alpha_{k,3}^{\dagger}(p+1) \beta_{k,3}(p)+\mathrm{H.C.}\right] \\
	&-t \sum_{k}\left[\sum_{ p=3 \atop  m\in \text {odd}}^{N-2} v_2\beta_{k,2}^{\dagger}(p) \alpha_{k,3}(p+1)+\sum_{  p=3 \atop  m\in \text {odd}}^{N-2} v_3\beta_{k,3}^{\dagger}(p+1) \alpha_{k,3}(p)+\sum_{  p=4 \atop  m\in \text {even}}^{N-3}v_1 \beta_{k,3}^{\dagger}(p) \alpha_{k,3}(p)+\mathrm{H.C.}\right], 
	\end{aligned}
	\label{x}
	\end{equation}
\end{widetext}
\noindent
where $\alpha_{k,q}(p)$ and $\beta_{k,q}(p)$ are the Fourier transformed operators, 
and $v_1 = e^{-ika_T/9}$, $v_2 = e^{ika_T/18}$, and $v_3 = e^{-ika_T/18}$, with  
$a_T = 1$ the unit cell size. 

\section{Tight-binding with next-nearest-neighbor hopping}\label{ap:seca2}

\begin{figure}[ht]
\includegraphics[width=0.5\textwidth]{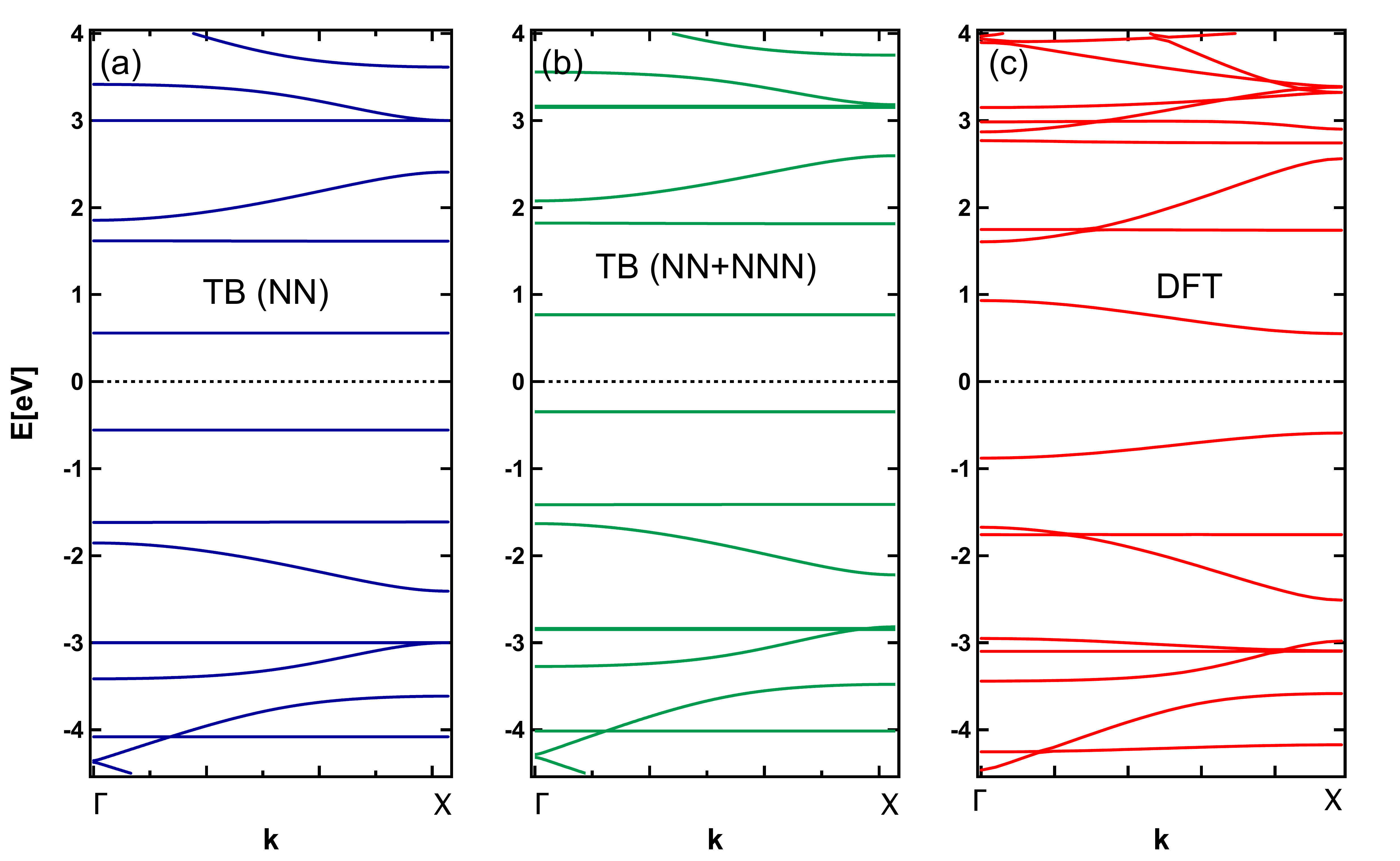}
	\caption{(a) Tight-binding band structure for 3-AGNR(1,3) with 
	NN hoppings only [reproduced from Fig.~\ref{fig10}(b)]. (b) Same as in panel 
	(a), but adding a NNN hopping $t_{NNN}$ 
	to the calculations, with $t_{NNN}=0.1$~eV.~\cite{Neto2009} (c) DFT results 
	for 3-AGNR(1,3) [reproduced from Fig.~\ref{fig10}(a)]. 
}
\label{figA2}
\end{figure}

In Fig.~\ref{figA2} we present tight-binding and DFT results 
to asses the stability of the tight-binding flat-bands to the 
addition of a NNN hopping $t_{NNN}$ to the calculations. 
In Fig.~\ref{figA2}(a) we reproduce the tight-binding bands shown previously in 
Fig.~\ref{fig10}(b) for 3-AGNR(1,3), which included just nearest-neighbor (NN) hoppings. 
In Fig.~\ref{figA2}(b) we add NNN hoppings $t_{NNN}=0.1$~eV~\cite{Neto2009} to the 
calculations. As expected, the results are not particle-hole symmetric anymore. 
However, all the flat-bands (in the interval of energy shown) remain flat. 
Thus, since the DFT results [in panel (c), reproduced from Fig.~\ref{fig10}(a)] 
show that flat-band $1$ (the closest to the Fermi energy) has acquired 
dispersion, we conclude that longer hoppings than NNN are 
necessary in the tight-binding calculations to produce dispersion in flat-band $1$. This 
can be understood by looking at the Wannier-like state for this band, shown in 
Fig.~\ref{fig4}(a). There, we clearly see that, to connect two unit cells, it 
is necessary at least a 3$^{\rm rd}$ NN hopping. This may explain too, why flat-band 
$3b$ has acquired a small dispersion, while flat-band $4$ has acquired just a 
slight dispersion. 

\section{Results for `Staggered' heterostructures}\label{ap:seca3}

\begin{figure}[ht]
\includegraphics[width=0.5\textwidth]{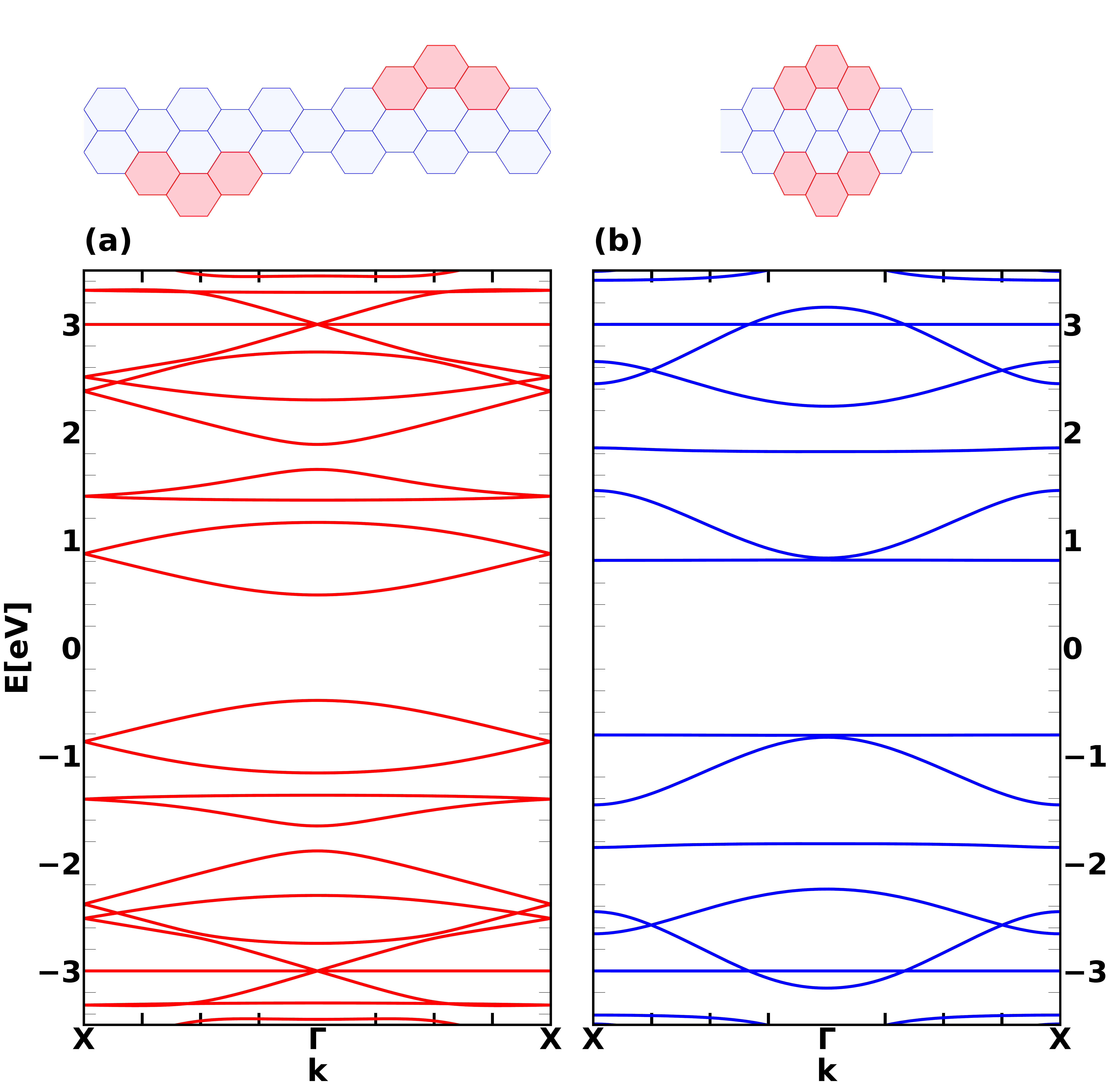}
	\caption{(a) Tight-binding band structure for 5-AGNR-S(1,3). 
	(b) Tight-binding band structure for 5-AGNR(1,3). The unit 
	cell of each heterostructure is shown at the top of each panel. 
}
\label{figA3}
\end{figure}

In Refs.~\cite{Groning2018,Rizzo2018} a second type of 
heterostructure has been introduced, less symmetric than the one we 
analyzed in this work. The reason we did not focus our attention 
in these so-called `Staggered' heterostructures is that they show 
less flat-bands than the so-called `Inline' heterostructures 
(which were the focus of this work). To exemplify that, in Fig.~\ref{figA3}(a) 
we compare the tight-binding band structure results for a 5-AGNR-\textbf{S}(1,3) 
heterostructure [panel (a)] with that for a 5-AGNR-(1,3) one [panel (b)]. 
Notice the inclusion of an `\textbf{S}' (in bold, for \textbf{S}taggered) 
to the label for the heterostructure. On top of Fig.~\ref{figA3}(a) we show 
a single unit cell for the 5-AGNR-\textbf{S}(1,3) heterostructure. By comparing 
it to the single unit cell on top of panel (b) [for 5-AGNR-(1,3)], which was 
described in Sec.~\ref{sec:model}A, it is easy to understand the meaning of 
the (1,3) nomenclature, since the idea is the same as the one introduce in 
Sec.~\ref{sec:model}A. 

By comparing the two panels, one notices that only the $\pm t$ flat-bands 
have survived in the Staggered heterostructure. 
We have checked that what appears to be two 
flat-bands (touched by a dispersive band, located, respectively,  
between energies $-1$~eV and $-2$~eV, and below energy $-3$~eV) are in 
reality two slightly dispersive bands, and not perfectly flat, like the $\pm t$ flat-bands. 
Our conclusion also rests in the fact that we could not discern a clear 
Wannier-like state associated to them. Thus, in the 5-AGNR-S(1,3) heterostructure 
there are just $\nicefrac{1}{4}$ of the flat-bands present in the (Inline) 5-AGNR(1,3) 
heterostructure, shown in panel (b). 

\begin{figure}[ht]
\includegraphics[width=0.5\textwidth]{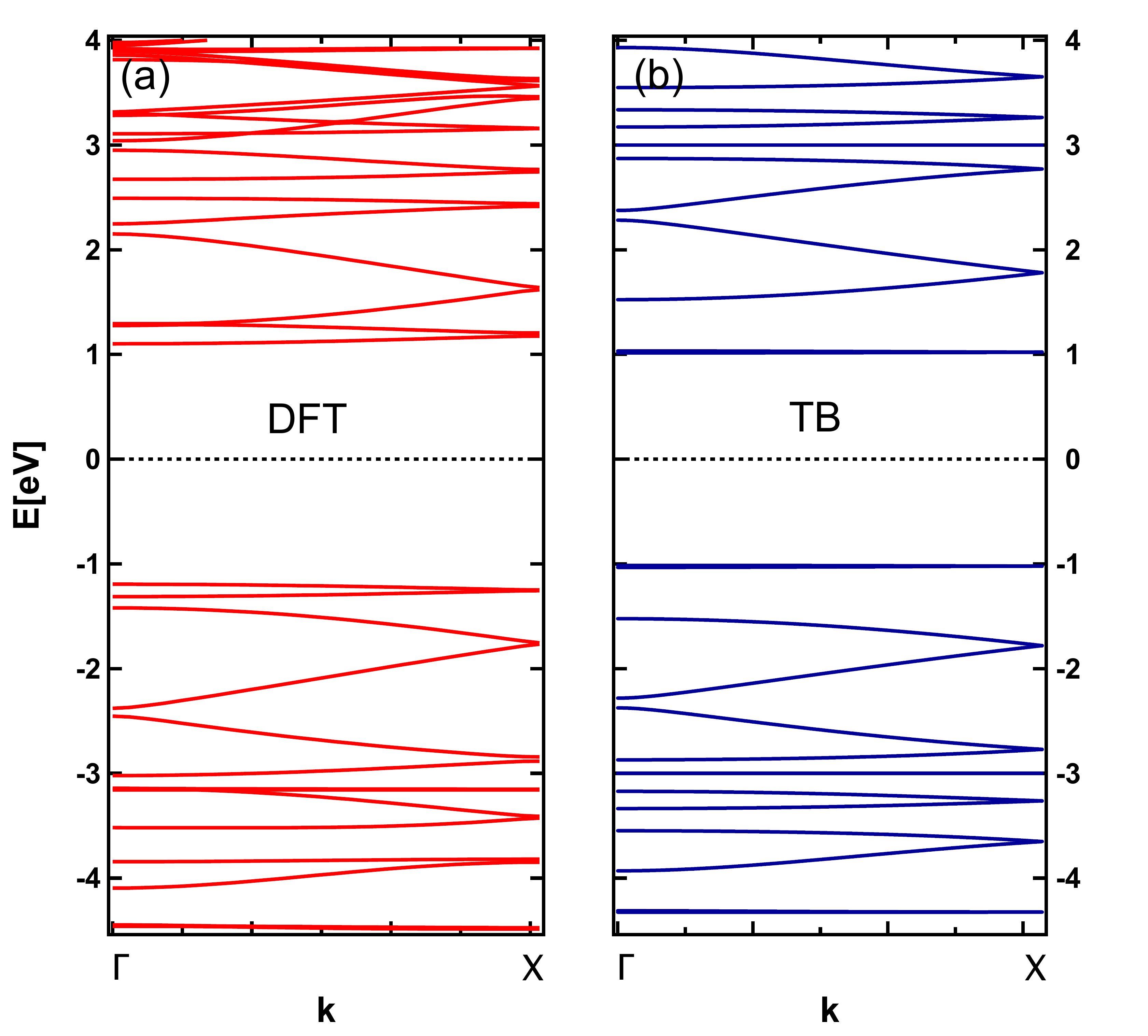}
	\caption{(a) DFT band structure for 3-AGNR-S(1,3). 
	(b) Tight-binding band structure for 3-AGNR-S(1,3). 
}
\label{figA4}
\end{figure}

Finally, for completeness sake, in Fig.~\ref{figA4} we show a 
comparison of the DFT band structure for 3-AGNR-S(1,3), with the tight-binding 
band structure, in panels (a) and (b), respectively. Aside from the expected broken particle-hole 
symmetry in the DFT bands, it is easy to see the very good agreement 
between the two results. A careful analysis of the DFT results shows that the 
only flat-band that is perfectly non-dispersive is the $\pm t$ flat-band 
(located just below $-3$~eV), reinforcing our claim that the Inline 
heterostructures have more robust flat-bands. 

\bibliography{ssh-flat}

\begin{thebibliography}{76}
\bibitem [{not()}]{note0}%
  \BibitemOpen
  \href@noop {} {}\bibinfo {note} {Note that, to improve readability, we show
  just the negative energies. In addition, to facilitate comparison with
  Fig.~\ref{fig2}, in panel (a) we label bands $1$ to $4$ accordingly, although
  they have obviously changed their positions (except for band $3$) in relation
  to Fig.~\ref{fig2}.}\BibitemShut {Stop}%
\end{thebibliography}

\end{document}